\documentclass[10pt]{article}

\usepackage{amsmath}
\usepackage{amssymb}

\usepackage{graphicx}

\usepackage{cite}

\usepackage{color} 
\usepackage[abs]{overpic}
\usepackage{dcolumn}
\usepackage{bm}
\usepackage{subfigure}
\usepackage{subeqnarray}

\usepackage[labelfont=bf,labelsep=period,justification=raggedright]{caption}

\makeatletter
\renewcommand{\@biblabel}[1]{\quad#1.}
\makeatother

\date{}

\begin{document}

\begin{flushleft}
{\Large
\textbf{Epidemic spreading on preferred degree adaptive networks}
}
\\
Shivakumar Jolad$^{1,2,\ast}$, 
Wenjia Liu$^{1,3}$, 
B. Schmittmann$^{1,3}$
R. K. P. Zia$^{1,3}$
\\
\bf{1} Department of Physics, Virginia Tech, Blacksburg, VA , USA \\
\bf{2} Indian Institute of Technology - Gandhinagar, Ahmedabad, Gujarat , India \\
\bf{3} Department of Physics and Astronomy, Iowa State University, Ames, IA , USA 
\\
$\ast$ E-mail: Corresponding shiva.jolad@iitgn.ac.in
\end{flushleft}


\section*{Abstract}
We study the standard SIS model of epidemic spreading on networks where
individuals have a fluctuating number of connections around a preferred
degree $\kappa $. Using very simple rules for forming such preferred degree
networks, we find some unusual statistical properties not found in familiar
Erd\H{o}s-R\'{e}nyi or scale free networks. By letting $\kappa $ depend on
the fraction of infected individuals, we model the behavioral changes in
response to how the extent of the epidemic is perceived. In our models, the
behavioral adaptations can be either `blind' or `selective' -- depending on
whether a node adapts by cutting or adding links to randomly chosen partners
or selectively, based on the state of the partner. For a frozen preferred
network, we find that the infection threshold follows the heterogeneous mean
field result $\lambda _{c}/\mu =\langle k\rangle /\langle k^{2}\rangle $ and
the phase diagram matches the predictions of the annealed adjacency matrix
(AAM) approach. With `blind' adaptations, although the epidemic threshold
remains unchanged, the infection level is substantially affected, depending
on the details of the adaptation. The `selective' adaptive SIS models are
most interesting. Both the threshold and the level of infection changes,
controlled not only by how the adaptations are implemented but also how
often the nodes cut/add links (compared to the time scales of the epidemic
spreading). A simple mean field theory is presented for the selective adaptations which
capture the qualitative and some of the quantitative features of the
infection phase diagram.

\section*{Introduction}

Concepts and tools from network science provide a powerful framework for the
description of many physical, biological, and social systems, from the world
wide web to neural architectures and from Facebook to power grids \cite%
{AlbertB.RMP.2002, Estrada.Springer.2010}. In the initial years of the
growth of network science, researchers focused on characterizing the network
topology, \cite{Watts.Nature.1998,AlbertB.RMP.2002} and then studying the
time-dependent processes on complex static networks \cite%
{BarratBV.Cambridge.2008, DorogovtsevGM.RMP.2008}. Often the
\textquotedblleft dynamics \emph{on} networks\textquotedblright\ was treated
distinctly from the `dynamics \emph{of} networks.\textquotedblright\ However
many recent studies have focused on more realistic situations where dynamics 
\emph{of} the network and dynamics \emph{on} the network are coupled
together, with a non-trivial feedback loop connecting them \cite%
{GrossDLimaB.PRL.2006,Gross.JRSIn.2008}. In this work, we study the
spreading of infectious diseases on a network of interpersonal connections
where the adaptive behaviors of the affected population influence both the
disease dynamics and the network topology.

The behavior of classic epidemic models such as
susceptible-infected-susceptible (SIS) model and the
susceptible-infected-recovered (SIR) model \cite%
{AndersonMay.Oxford.1992,Daley.Cambridge.2001} has been widely studied on
regular lattices and on specific networks such as random, small world or
scale-free networks \cite%
{MooreN.PRE.2000,Pastor-SatorrasV.PRE.2001,BarratBV.Cambridge.2008} (see 
\cite{Keeling.JRSi.2005} for review). These studies assume that the disease
spreads on a \textit{static} network with characteristics which are \textit{%
independent }of the nodes. However, in a dynamic social setting, people are
likely to respond by social distancing or quarantine -- changes in behavior
that are perceived to reduce the likelihood of infection. Such behavioral
adaptations will change the network topology and feed back into the dynamics
of epidemic spreading. Recently, there has been growing interest to include
such adaptive behavior in epidemic models. Given the wide range of human
responses and their impact on the spread of the disease, modeling all these
possibilities seems difficult and daunting. Thus, it is natural to consider
simplified models with a few effective parameters. While such models cannot
predict the epidemiological or social details quantitatively, they may be
able to provide insight into qualitative and universal features of how
adaptive behavior impacts the dynamics of epidemics. In this spirit, we
introduce our models and study their properties.

Funk \textit{et al} \cite{FunkSJ.JRSI.2010} classify the current literature
on adaptive epidemic models based on the source of information (local or
global) and the type of information (belief or prevalence) about the
epidemic. Belief-based models emphasize individuals' awareness of a disease,
and how they evaluate the associated dangers \cite{EpsteinPCH.PlosOne.2008,
FunkGWJ.PNAS.2009, TanakaKF.ThPopBio.2002, KissCRS.MathBio.2010}. For
example, some authors have modeled risk perception by decreasing the
infection rate with the fraction of infected individuals in the local
network of the node \cite{BagnoliLS.PRE.2007} and by introducing voluntary
vaccinations \cite{PerisicB.PlosCompBio.2009, BauchGE.PNAS.2003}.
Prevalence-based models emphasize the objective assessment of the extent of
epidemic spread and personal risk. Most of these studies have concentrated
on coupling disease dynamics with network adaptations through rewiring of
links \cite%
{GrossDLimaB.PRL.2006,Gross.JRSIn.2008,WangCSA.JPhysA.2011,ZanetteR.JBioPhys.2008,MarceauPHAD.PRE.2010,GrossK.EPL.2008,SchwarzkopfRM.PRE.2010,ShawS.PRE.2008}
and studying the dynamics of S-I, S-S and I-I links. 
One might argue, however, that such rewiring models make a somewhat
unrealistic assumption, namely, that individuals necessarily create a link
with a healthy person after cutting a link with an infected one.

We address some of the limitations of prevalence-based epidemic studies by
proposing a new type of network which contains a natural parameter, $\kappa $%
, the `preferred degree.' An individual (a node) with more/fewer contacts
than $\kappa $ will tend to cut/add links. This parameter allows us to
easily model \emph{adaptive} behavior depending on the (perceived) level of
threat from an epidemic. Let us point out several other advantages of this
approach. Our network does not have unrealistically large degrees
responsible for epidemics with vanishing thresholds \cite%
{CastellanoP.PRL.2010}. Our model can easily be generalized to endow
different nodes with different $\kappa $'s, e.g., to account for the
presence of extroverts and introverts\cite{PlatiniZ.JStat.2010,
LJSZ.TBS.2011} in our society. Recent work has attempted to synthesize
more realistic network such as those based on survey and census data \cite%
{Eubank.Nature.2004,Mossong.PLoS.2008},  and trajectories of mobile phone
users \cite{Gonzalez.Nature.2008}. Models based on
realistic features of social network such as assortativity (homophily) \cite%
{Mcpherson.ARSoc.2001} in social networks and range of interactions (like
close and casual) have received considerable attention \cite%
{Estrada.PRE.2011}. Our network model can be used to simulate features of
these `realistic' networks by making preferred degree distribution match the
`true' distribution and tuning the clustering coefficient by methods such as
the one developed by Volz \cite{Volz.PRE.2004}. 

 We highlight few major differences between our approach and the literature on prevalence and global information based adaptive networks. In the rewiring approach  \cite{GrossDLimaB.PRL.2006,Gross.JRSIn.2008}, the total number of links in the population is \textit{fixed for all time}, regardless of the level of the epidemic. By having a preferred-degree (which adapts to the state of the epidemic), the total number of contacts in the population is reduced when the disease spreads dramatically and returns to "normal" levels when the epidemic recedes.  In this sense, our adaptive preferred degree plays a role analogous to the rewiring rate, in delaying the onset of an epidemic.  Zanette and  Risau-Gusm\'an \cite{ZanetteR.JBioPhys.2008} consider case where susceptible agents can decide to break links with their infected peers and links are permanently broken. In our approach, no link is permanently broken as the dynamics is kept active by infected nodes who can reconnect with any susceptible.

We begin by modeling the simplest case, where all nodes are characterized by
a single $\kappa $, i.e., a homogeneous population. The network is dynamic,
so that nodes can add or delete links, in an attempt to reach or maintain $%
\kappa $. When a disease spreads on this network, the detailed dynamics of
adding/cutting links changes in response to the epidemic. In the following,
we propose a model reflecting \emph{global prevalence-based} information, by
letting $\kappa $ depend only on $\phi $, the fraction of infected
individuals in the entire population. We model two typical human response:
(a) If individuals are not aware who is infected and who is healthy (an
`invisible' disease, e.g., AIDS), they may cut (or add) links blindly in
response to news of a raging epidemic. We will refer to this adaptive
behavior as `blind response.' (b) If the disease is `visible' (e.g., the
flu), an individual is more likely to be more discriminating when cutting or
adding a contact -- a response we naturally label as `selective.' Here, the
dynamics of network will depend on the state of the recipient node:
Susceptible individuals will preferentially cut links with the infected and
add links with other susceptibles. For the \textit{blind}
adaptations, we investigate three types of behavior: the \textit{reckless}
(where $\kappa $ remains constant, then drops abruptly only when $\phi $
reaches some large value), the \textit{typical} (where $\kappa $ decreases
linearly with increasing $\phi $, leveling off at some constant $\kappa
_{\min }$), and the \textit{nosophobic} (who cut ties precipitously as soon
as $\phi $ deviates from zero). We find that the epidemic threshold does not
change, but the level of epidemic depends on the `degree of fear' in the
population. For the \textit{selective} adaptations, we focus only on the
reckless and typical types. Here, both the threshold and the level of
infection change. We develop a mean field theory for local adaptations by
writing equations for node and link dynamics. The predictions of this theory
predict all the qualitative features of the simulations.

Our paper is organized as follows: In section I, we set the scene:
presenting the formation of preferred degree networks and introducing an SIS
dynamics on this network (initially with no adaptive features). We will
summarize two theoretical approaches: a simple mean field theory (MF) and more sophisticated annealed adjacency matrix (AAM) method \cite{GuerraG.PRE.2010}. We also compare our results for the critical $%
\lambda _{c}$ with predictions of heterogeneous mean field theory \cite{Pastor-SatorrasV.PRL.2001,MorenoPV.EPL.2002} . In section II, we turn to study populations with adaptive response to a raging epidemic. In section III, we describe our main results for adaptive epidemic propagation. Section III.a deals with blind adaptations where  a given nodes cannot ``see'' the disease states of the connected nodes.. The SIS phase diagram and degree distribution for these adaptive cases are much richer than those in non-adaptive networks. Much of the phase diagram is captured quite well by a simple mean field theory. In section III.b, we discuss the cases with selective adaptations. Simulation results are compared with a mean field
theory, the details of which can be found in appendix (see supplementary information). We conclude, in the last section, with a discussion of our results and their implications for future research.

\section*{Analysis}

\textbf{\large \\ I ~~ SIS on preferred degree networks}


\textbf{ \\ I.a ~~ Network formation \\}

\begin{figure*}[tbp]
\begin{center}
\includegraphics[scale=0.35]{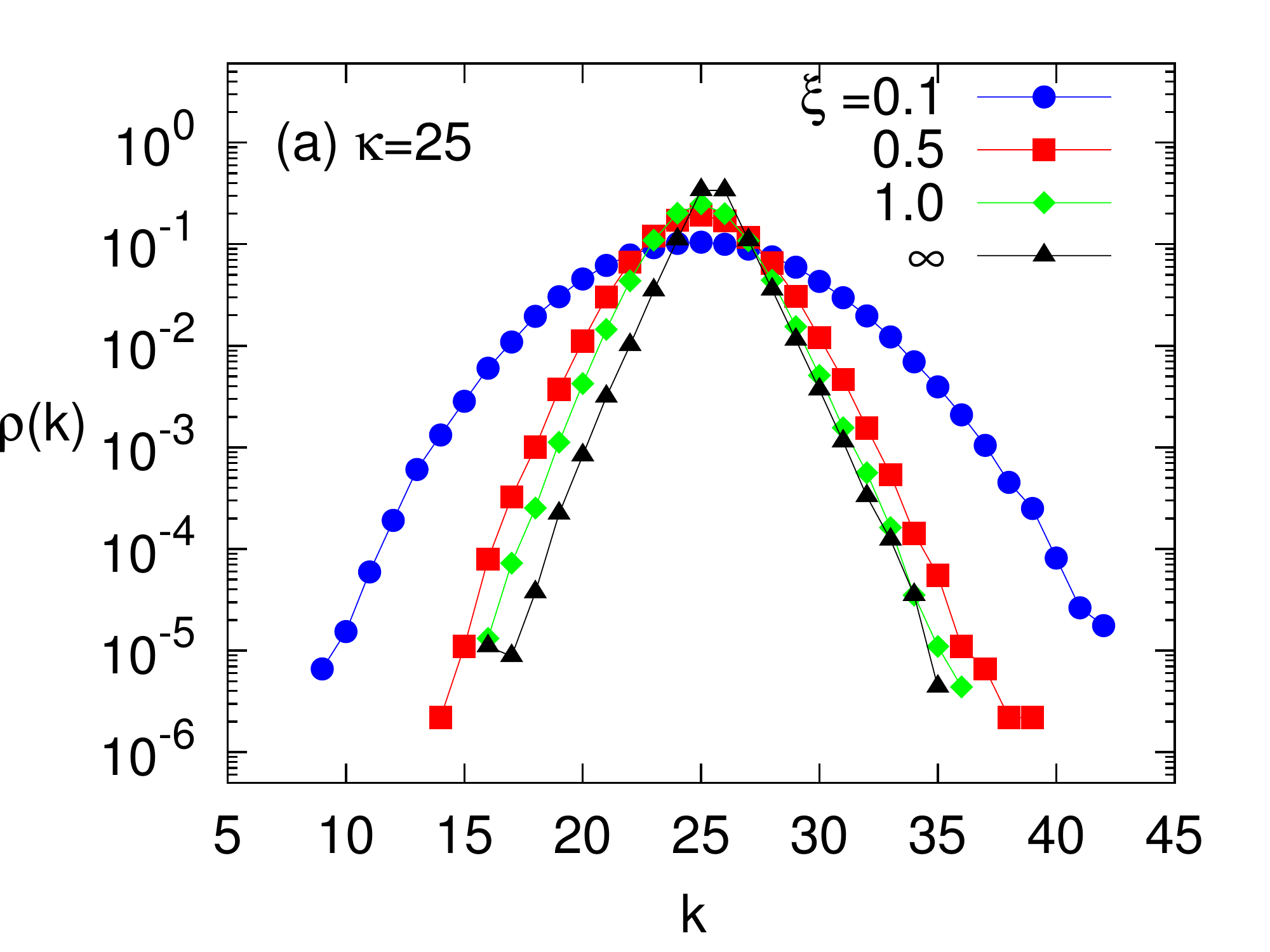} %
\includegraphics[scale=0.35]{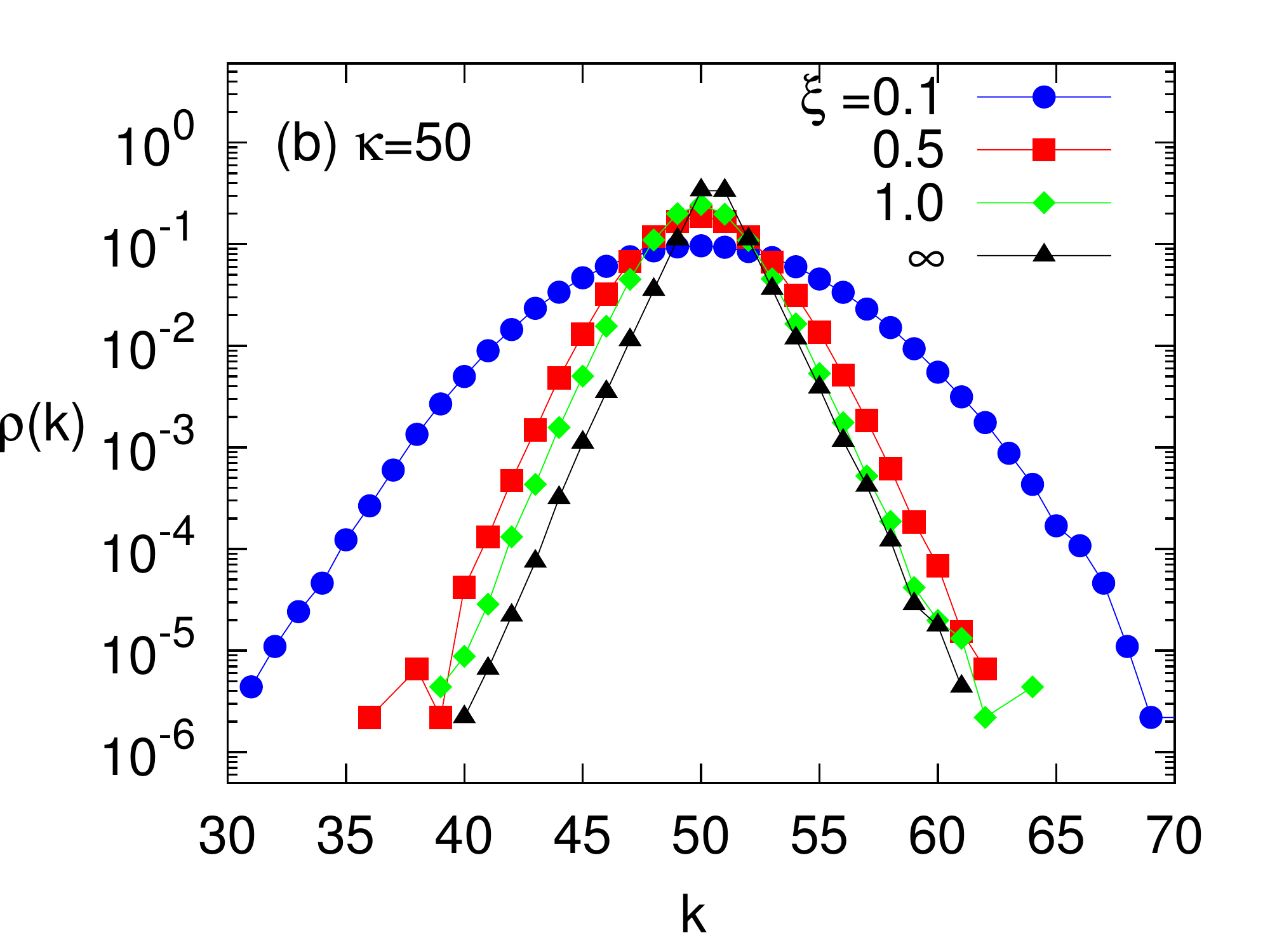}
\end{center}
\par
\vspace{-0.5cm}
\caption{[Color online]{\bf Degree distribution of preferred degree networks.} 
Networks with  $N=5000$ nodes and various inflexibility parameters $\xi$ (see Eq. \ref{plink}). Panels (a) and (b) corresponds to $\kappa=25$  and $\kappa=50$ respectively. The $\xi\rightarrow \infty $ corresponds to
the totally inflexible individuals and results in a Laplace distribution.}
\label{Fig1}
\end{figure*}

To explore the behavior of epidemics on dynamic networks, let us first
present the foundation, i.e., a network with preferred degree(s). Following
the lines introduced in \cite{PlatiniZ.JStat.2010}, we briefly review how
such a network is formed and evolves. Details of the statistical properties
of such networks are also of interest, but will be presented in another
publication \cite{LJSZ.TBS.2011}. For simplicity, we first consider a
homogeneous population, i.e., a system with $N$ nodes (individuals) of
identical behavior, evolving stochastically. In each time step, a random
node $n$ ($=1,2,\cdots ,N$) is selected and its degree, $k_{n}$, is noted.
Then, an attempt to add (cut) a link is made,  with probability $w_{+}(k_{n})$
 $\left (w_{-}(k_{n})\right)$ . Although an infinite variety of $w_{\pm }$'s is possible,
we impose some general properties which mimic typical human behavior, e.g., $%
w_{+}(0)\cong 1$ and $w_{-}(k\gg 1)\rightarrow 1$, as well as the logical
constraint $w_{-}(0)=0$. A simple choice, used in all our simulations, is $%
w_{-}(k)=1-w_{+}(k)$, with
\begin{equation}
w_{+}(k)=\frac{1+e^{-\kappa \xi }}{1+e^{(k-\kappa )\xi }},  \label{plink}
\end{equation}%
recognizable as a Fermi-Dirac function. Here, $\xi $ plays the role of
`inflexibility' (or `rigidity') of the personality, so that a node
(individual) with $\xi =\infty $ will always cut/add a link when it finds
itself with more/fewer links than $\kappa $. Indeed, apart from a brief
digression in the next paragraph, the step function is used in all the
simulations presented here. In the code, we choose $\kappa $ to be slightly
larger than an integer, so that a node with $k\leq \kappa $ will attempt to
add a link. Note also that, with $w_{-}=1-w_{+}$, the network will always
change, by the addition or deletion of a link. The partner node for this
action is randomly chosen out of the eligible pool. Thus, the `recipient'
has no control over a link to it, whether created or destroyed. In a Monte
Carlo step (MCS), $N$ such attempts are made, so that there is one chance,
on the average, for each node to add or cut a link.

With a preferred degree, our network is clearly not scale-free. Also, unlike
the case of a Erd\H{o}s-R\'{e}nyi network, the degree distribution in the
steady state here, $\rho (k)$, is \textit{not Gaussian}. Though $\rho $
depends on the details of $w_{\pm }$, we discover a universal feature:
exponential tails when $k$ is far from $\kappa $. In Figure. \ref{Fig1}, we
show typical simulation results for $\rho $ (with $N=5000,\kappa =25,50$).
Indeed, for a group of completely rigid individuals ($\xi \rightarrow \infty 
$), $\rho (k)$ is a Laplace distribution ($\propto e^{-\ln 3|k-\kappa |}$).
With a more flexible group ($\xi \leq 1$), the maximum around $\kappa $ is
rounded off, up to a width of $\approx 1/\xi $, before crossing over to the
same kind of exponential tails. This behavior is heuristically understood in
the context of an approximate master equation, details of which can be found
elsewhere \cite{ZiaLJS.PhysProc.2011,LJSZ.TBS.2011}. Our main focus in the
remainder of this article will be the SIS dynamics associated with the
nodes, evolving along with this changing network.

 \textbf{ \\ I.b ~~ SIS on static and dynamic  preferred degree networks \\}

Having presented the dynamics of a network with static nodes, we now endow
the nodes with their own degrees of freedom. Following the standard SIS
model \cite{AndersonMay.Oxford.1992}, we assign a binary state variable, $%
\sigma _{n}=0,1$, to node $n$, corresponding to that individual being
susceptible ($S$) or infected ($I$). The system evolves by discrete attempts
to update a randomly chosen node. If it is infected, then it recovers with
rate $\mu : I\xrightarrow{\mu}S$. If it is susceptible, then the
disease is transmitted with rate $\lambda $ from each of its infected
contacts $S+I\xrightarrow{\lambda}I+I$ ( Here we set the time step equal to 1 making rates same as probabilities).
We consider infection as a simultaneous event, so that an $S$ in contact with $m$ infected nodes will
contract the disease with probability $1-\left( 1-\lambda \right) ^{m}$ ($%
\rightarrow m\lambda $ if $\lambda \ll 1$). Again, a MCS is defined as $N$
such attempts.

A good measure of the `level of the epidemic' is the fraction of infected
nodes: $\phi \equiv \Sigma _{n}\sigma _{n}/N$. Clearly, a population with $%
\phi =0$ will not evolve, a state known as `absorbing.' If the initial state
has $\phi >0$, then the epidemic may die out (i.e., $\phi \rightarrow 0$)
quickly or only over very long times, since there is a non-vanishing
probability ($\thicksim e^{-N}$) for a fluctuation to drop $\phi $ to $0$.
In the latter, known as an `active state,' $\phi $ is typically positive,
meaning that the epidemic is typically \textquotedblleft alive and
well.\textquotedblright\ Whether the system becomes active or not will
depend on network topology and the ratio $\lambda /\mu $. For simplicity, we
fix $\mu =0.5$ in all our simulations and use $\lambda $ as a control
variable. The goal is a phase diagram: Given $\lambda $ and a particular
network, will the epidemic die or stay active? and where is its threshold: $%
\lambda _{c}$?

While a well-defined set of such questions can be formulated for infinite
systems running for indefinite times, the task is less simple when
confronted with simulations with finite systems and finite run times. In
particular, since our systems will reach absorbing states in finite time, it
is difficult to pin point the threshold, near which the typical $\phi $ is
vanishingly small. To overcome this difficulty, we introduced a trick into
our simulations. To prevent our system from falling into the absorbing
state, we do not allow the last $I$ to recover. We refer to such a node as an `immortal'. We stress that we do not fix a single node as immortal, but simply prevent the last infected node from recovering. 
The advantage of this approach is clear: Our system never ceases to
evolve, so that time averages in a steady state can be used to study
ensemble averages (both denoted by $\left\langle \cdot \right\rangle $). Of
course, we should keep in mind that, in the `inactive state,' $\left\langle
\phi \right\rangle \neq 0$ but $\mathcal{O}\left( 1/N\right) $. Further
measurements can be implemented to characterize this state in more detail.
For example, distributions of $\phi $ are expected to be exponential\ ($%
e^{-c\phi }$) and how $c$ varies with $\lambda $ should be revealing.

We first studied \emph{static }networks with a preferred degree, to provide
a baseline for later investigations with co-evolving networks. For this
study, we generated 50 network realizations using the scheme specified above
(using 10K MCS for each run) and kept them quenched as we continued with the
evolution of the nodes. After thermalization for 1000 MCS, we measure $\phi $
every 10MCS and then averaged over the 50 networks. The results for this
(quenched) average $\phi $, as a function of $\lambda $, display a clear
signal of the expected transition from inactive to active regimes of the
epidemic. Away from $\lambda _{c}$, the fluctuations over a run are about
1\%. The averages from the 50 realizations also do not differ by more than
this amount. Not surprisingly, close to the transition, fluctuations are
more substantial ($\thicksim 10\%$). Exploring the critical region
quantitatively is a worthwhile pursuit, but beyond the scope of this study.

\begin{figure}[tbp]
\centering
\begin{overpic}[scale=.4, tics=5,unit=1mm]{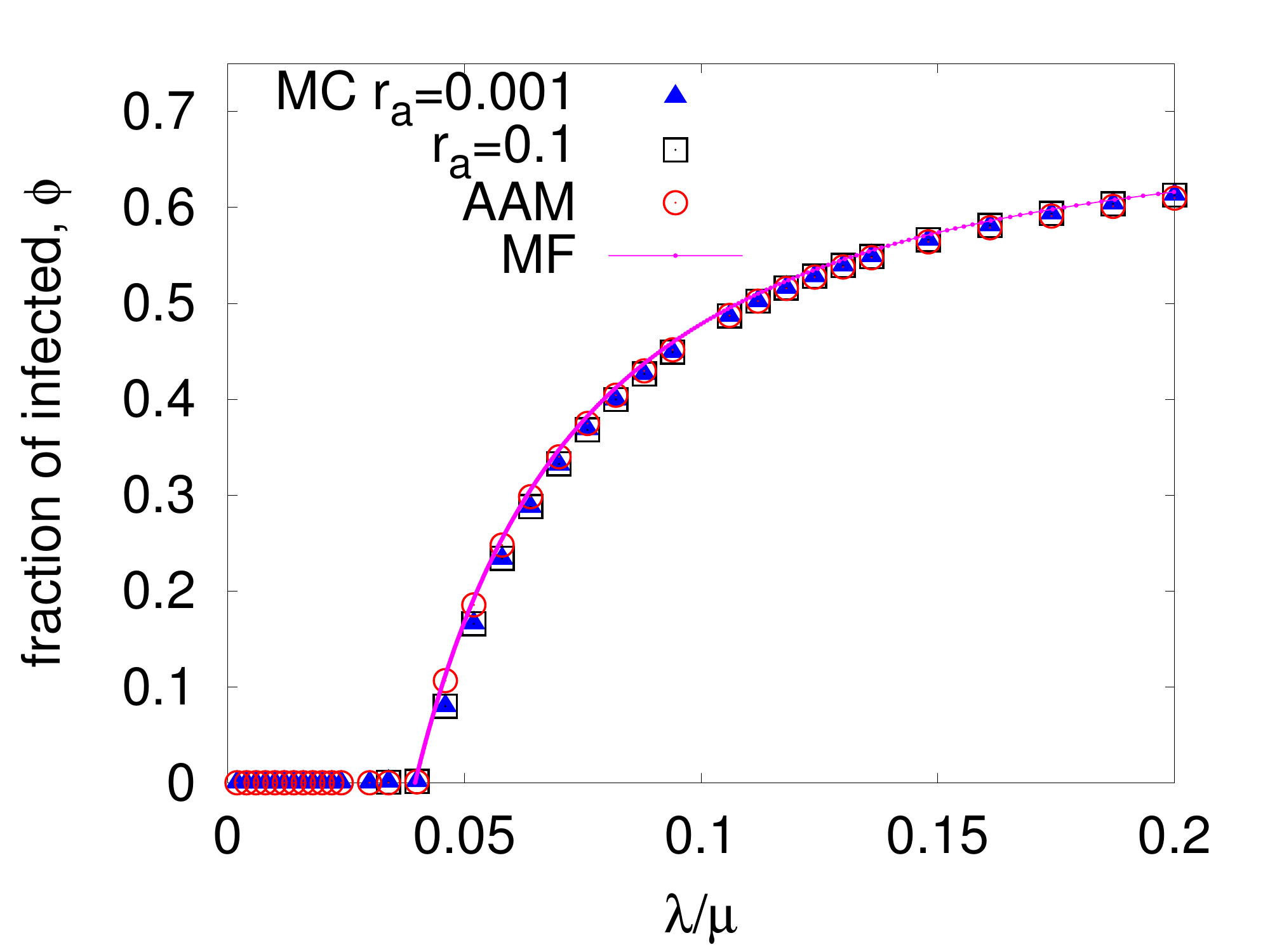}
\put(40,12){\includegraphics[scale=.18]{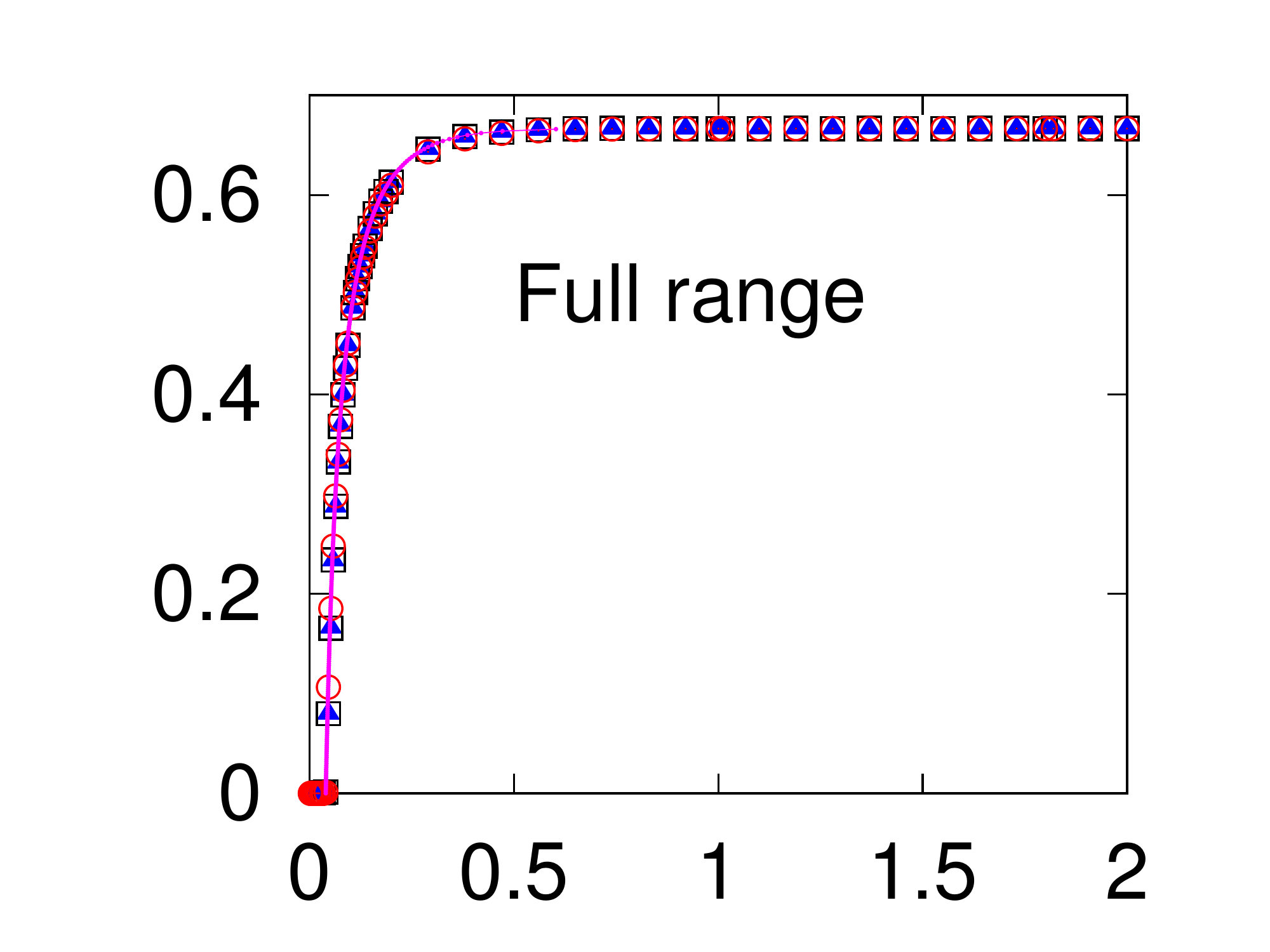}}
\end{overpic} 
\caption{[Color online] {\bf The SIS phase diagram for non-adaptive network.}
Fraction of infected population versus relative infection rate is plotted  the vicinity of the transition point $\lambda_c/\mu=0.04$ and compared with mean-field theories, for $N=5000,~\kappa=25,~\mu=0.5$, and two values of $r_a$. The numerically integrated AAM equations ( Eq 1. in \cite{GuerraG.PRE.2010}) are shown as open circles (red online), and results from the simple mean-field theory of Eq. \ref{phi-lambda} are plotted as solid lines (magenta online). }
\label{Fig2}
\end{figure}

Next, we turn our attention to SIS on \textit{dynamic} networks, where we
must account for the fact that network and disease dynamics typically
proceed at different time scales in society. Given that we are modeling the
former as a \textit{response} to a spreading epidemic, we will assume that
network timescales are slower.  In this spirit we choose the epidemic spreading to be $1/r_a$ ($r_a<1$) times faster than the network adaptations.
That is, for every one MC step of the network, we perform $1/r_a$ MCS of
nodes. Mostly, we use $r_a =0.1$. The SIS dynamics on a static network
consists of letting $r_a\rightarrow 0 $. In practice, we performed runs
with $r_a=0.001$ and found that $\phi \left( \lambda \right) $ is not very
sensitive to $r_a$ and that the $r_a=0.001$ data are indistinguishable from
those in static networks above. In Figure. \ref{Fig2}, we present results from
runs with $r_a=0.1$ (open black squares) and $r_a=0.001$ (solid blue
triangles), leading us to the conclusion that, within our statistical
errors, the time scales of network dynamics have little effect on an
epidemic in a homogeneous population.  We point to the readers that we present the results for time averaged data. Detailed investigations into the fluctuating dynamics is beyond the scope of the present work.   For a recent work on instantaneous time description of network dynamics we refer the reader to   \cite{PerraGPV.SciRep.2012}.  In the next subsection, we will present theoretical perspectives of this system and how such phenomena can be understood.

\textbf{\\ I.c ~~ Simple mean field theory and the annealed adjacency matrix approach \\}

To attack a statistical system theoretically, the first and simplest tool is
a mean field (MF) approach. Since our interest is the long time behavior of $%
\phi _{t}\left( \lambda ,\mu \right) $, this first step consists of writing
a simple equation for the evolution of $\phi _{t}$. Following standard MF
analysis, we write 
\begin{equation}
\frac{d\phi _{t}}{dt}=-\mu \phi _{t}+\left( 1-r\left( \phi _{t}\right)
\right) \left( 1-\phi _{t}\right) \,\,,  \label{MF-infection}
\end{equation}%
where the first term models the $I$'s recovering. In the second term, $%
r\left( x\right) =\left( 1-\lambda \right) ^{\kappa x}\label{r}$ is the
probability that an $S$ is \emph{not} infected by any of its infected
contacts. By setting the derivative to zero in Eq. \ref{MF-infection}, we
find stationary solutions (fixed points): $\phi =\phi _{t\rightarrow \infty
} $. For small/large $\lambda $, the stable $\phi $ is zero/positive,
corresponding to the inactive/active state. The transition is predicted to
occur at%
\begin{equation}
\lambda _{c}^{MF}=1-e^{-\mu /\kappa }\,\,,  \label{MF crit-lambda}
\end{equation}%
which reduces, for $\mu \ll \kappa $, to an easily understandable result: $%
\lambda _{c}^{MF}\simeq \mu /\kappa $. In the active state, $\phi \left(
\lambda \right) $ is given by the solution to $\mu \phi =\left( 1-r\left(
\phi \right) \right) \left( 1-\phi \right) $. In other words, it is the
inverse of the explicit $\lambda \left( \phi \right) :$%
\begin{equation}
\lambda =1-\left\{ \frac{1-\left( 1+\mu \right) \phi ^{MF}}{1-\phi ^{MF}}%
\right\} ^{1/\kappa \phi ^{MF}}\,\,.  \label{phi-lambda}
\end{equation}%
The result is presented as the solid line (magenta on line) in Figure. \ref%
{Fig2} and shows that, while slightly higher than the simulation results, it
indeed captures the essentials of the epidemics. In the vicinity of
criticality, the exponent in $\phi ^{MF}\propto \left( \lambda -\lambda
_{c}\right) ^{\beta }$ takes the expected MF value $\beta ^{MF}=1$.

In a dynamic or a quenched random network, this approach may seem too
simplistic. In previous studies of SIS models on irregular, static networks,
better approximations have been developed. Examples include the \textit{%
heterogeneous} mean field (HMF) theory \cite%
{Pastor-SatorrasV.PRL.2001,MorenoPV.EPL.2002} and the annealed adjacency
matrix (AAM) approach \cite{GuerraG.PRE.2010}. The former takes into account
a distribution of degrees, such as $\rho \left( k\right) $ in our case, and
provides the critical threshold at $\lambda _{c}^{HMF}=\mu \langle k\rangle
/\langle k^{2}\rangle $, i.e., $\lambda _{c}^{HMF}=\left. \frac{\mu }{%
\langle k\rangle }\right/ \left\{ 1+\frac{\Delta k^{2}}{\langle k\rangle ^{2}%
}\right\} $. It has been widely applied, with considerable success, to study
critical dynamics on various networks. For our study here, we present in
Figure. \ref{Fig1} the few cases of $\rho \left( k\right) $ for the preferred
degree networks used, showing that $\langle k\rangle =\kappa $ as expected
and $\Delta k^{2}/\langle k\rangle ^{2}\lesssim 1\%$. Hence, the simple MF
prediction ($\lambda _{c}^{MF}\simeq \mu /\kappa $) is quite adequate.
Further, as our interest lies in the dominant behavior of the epidemic over
the entire phase diagram, rather than details of the transition, there is no
compelling need for using this complex method. As our network is dynamic,
the AAM method may provide better predictions. Let us briefly summarize this
approach \cite{GuerraG.PRE.2010} here. While the full dynamics involves a
fluctuating adjacency matrix, in the AAM, the elements $a_{nl}$ of the full fluctuating adjacency matrix are approximated by the probability that nodes n and l are connected. The infection probability of nodes are evolved through a discrete
Markov equation (Eq. 1 in ref. \cite{GuerraG.PRE.2010}). Steady state values
of infection probabilities are used to calculate $\phi ^{AAM}$. Applying
this technique to our problem, we find that $\phi ^{AAM}$ (red circles in
Figure. \ref{Fig2}) follows $\phi ^{MF}$ (magenta lines) quite closely at the
transition region. As for $\phi \left( \lambda \right) $ in higher $\lambda $%
's, we show only the static network data and $\phi ^{AAM}$ in the inset of
Figure. \ref{Fig2}. As expected, the infected fraction simply saturates at $%
\phi _{\max }=1/(1+\mu )$. Clearly, the agreement between simulation results
and all theoretical approaches is quite good. Thus, as a first step towards
understanding epidemics on more complex, adaptive networks, we will rely on
the simpler mean field theory.

\textbf{\large \\ II ~~ Adaptive response to a raging epidemic \\}

In the networks presented above, whether static or dynamic, the degree of
each node is effectively fixed in time ($\thicksim \kappa $ in our model).
However, when an epidemic is present, individuals are likely to exhibit
`social distancing' behavior, by cutting ties or reducing the number of
non-essential contacts (as documented in, e.g., \cite{Hatchett.PNAS.2007,Caley.JRSi.2008}).
Apart from being an inherently natural response, cutting ties may also occur due to externally imposed
public policies \cite{SARS.2003,Hatchett.PNAS.2007}. When the state of the
disease is not easily discernible (e.g., AIDS), one's response will be to
sever links blindly. On the other hand, if the disease is `visible' (e.g.,
the flu), one can be more selective, by cutting only contacts with the
infected. Such adaptive behaviors can be easily accommodated in our model by
letting $\kappa $ change, in response to the level of the infection. In this
work, we will study the effects on the epidemic due to both `blind' and
`selective' adaptations. In particular, we investigate infection levels, $%
\phi $, and degree distributions, $\rho \left( k\right) $, in the steady
states.

\subsection*{II.a ~~ Models of response}

\begin{figure}[tbp]
\centering
 \includegraphics[scale=0.40]{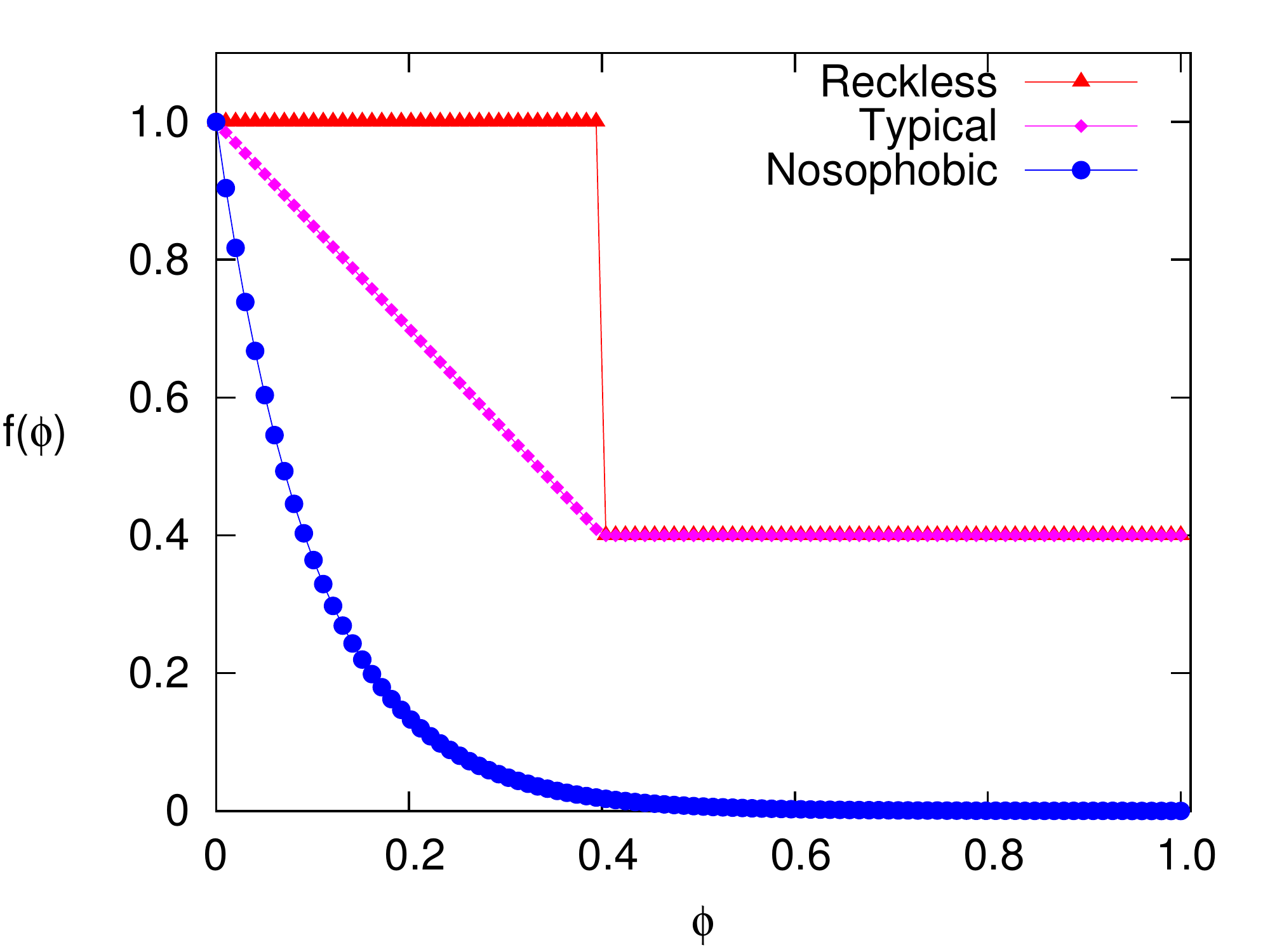}
\caption{[Color online] {\bf Adaptive fear factor.}
The ``fear factor'' $f(\phi )$ depending on the global infected fraction  $\phi$ (see Eq. 
\ref{fearfunction}) associated with different behavioral patterns
listed in section III.A . }
\label{Fig3}
\end{figure}

To incorporate adaptive behavior, our first task is to specify how the
population will lower the preferred degree, $\kappa $, in response to a
rising infection level. When an individual becomes aware of an epidemic, the
response is likely a combination of rational/prudent behavior and irrational
perceptions of the dangers. Though a typical population is diverse and
heterogeneous, we begin with the simplest system: a homogeneous population
with a unique response based on just one piece of information of the
epidemic, namely, the global infection level $\phi $. In other words, we let
every node update with the same $\kappa \left( \phi \right) $. For
convenience, $\kappa \left( \phi \right) $ is introduced via a `fear factor' 
$f\left( \phi \right) $: 
\begin{equation}
\kappa \left( \phi \right) =\kappa _{0}f\left( \phi \right) ;\quad f\left(
0\right) =1.  \label{fearfunction}
\end{equation}
Here, $\kappa _{0}$ is just the preferred degree for an uninfected
population, while $f$ is a monotonically decreasing function, which serves to reduce the preferred degree. Of the infinitely many behavioral patterns that can be modeled, we consider only three kinds here (Figure. \ref{Fig3}):

\begin{itemize}
\item \textbf{Reckless} individuals are oblivious to a low level of epidemic
present in the population. They keep the same $\kappa $ until the
epidemic reaches a certain threshold: $\phi _{\theta }$. (We assume $\phi
_{\theta }$ to be some fraction of $\phi _{\max }$.) At this point, they
abruptly change their preferred degree to $\kappa _{min}$. Keeping in mind
that a typical person would maintain a minimal set of contacts (family,
caretakers, etc.) even in the face of a raging epidemic, we simply choose $%
\kappa _{min}$ to be independent of $\phi $ for all levels higher than $\phi
_{\theta }$. Explicitly, $f_{reckless}(\phi )=\Theta \left( \phi _{\theta
}-\phi \right) +\left( \kappa _{min}/\kappa _{0}\right) \Theta \left( \phi
-\phi _{\theta }\right) $, where $\Theta $ is the Heaviside step function.
For simulations, we choose $\kappa _{0}=25,\kappa _{min}=10$, and $\phi
_{\theta }$ to be 60\% of the maximum $\phi _{\max }=1/\left( 1+\mu \right) $%
. Since we fix $\mu $ at $0.5$, we use $\phi _{\theta }=0.4$.

\item \textbf{Typical} individuals are likely to cut their contacts in a
more measured fashion. For them, we choose a linearly decreasing $f\left(
\phi \right) $. If this decrease is rapid enough, then these individuals'
comfort level would reach the lower limit ($\kappa _{min}$) before the
infection rate reaches its maximum level $\phi _{\max }$. Again for
simplicity, we let their $\kappa $ remain at $\kappa _{min}$ for all higher
levels of infection. Explicitly, $f_{typical}(\phi )=\left( 1-\alpha \phi
\right) \Theta \left( \phi _{\theta }-\phi \right) +\left( \kappa
_{min}/\kappa _{0}\right) \Theta \left( \phi -\phi _{\theta }\right) $,
where the slope and the threshold are related by $\alpha \phi _{\theta
}=1-\kappa _{min}/\kappa _{0}$. For this set of simulations, we chose the
same parameters as above: $\phi _{\theta }=0.4,\kappa _{0}=25,\kappa
_{min}=10$.

\item \textbf{Nosophobia} is an irrational fear of contracting diseases. To
model such a population, we let $f$ drop exponentially, as soon as the
slightest infection is detected. These individuals would eventually avoid
all personal contact. Explicitly, we have $f_{nosophobic}(\phi )=\exp (-\phi
/\phi _{s})$. With $\phi _{s}$ setting the severity of this phobia, we use $%
\phi _{s}=0.1$ in our simulations.
\end{itemize}

Of course, any real population will have a mix of these behaviors, with
perhaps time dependent compositions. Our hope is that studying these
homogeneous cases separately will help us untangle the effect of different
adaptive behavior on the epidemics.  To summarize
our model so far, when a node is chosen for updating its links, we measure
its degree $k$ and take note of the overall infection level ($\phi $). Then
we add/cut a link if $k$ is less/greater than $\kappa \left( \phi \right) $.
Choosing which link to add/cut and its affect on disease dynamics will be the focus of the next section.

\section*{Results}

\textbf{\large {III~~ Epidemic propagation in adaptive networks } }

\textbf{\\ III.a~~ Blind adaptation \\}

With an invisible disease, an individual does not know which of his/her
contacts (or potential contacts) is infected. As a result, adapting to the
news of say, a rising level of the epidemic, he/she simply cuts 
links to randomly chosen partners (as described in Section I) until a
smaller $\kappa \left( \phi \right) $ is reached. Similarly, if $k<\kappa
\left( \phi \right) $, the new contact will be also chosen blindly. Setting
aside the interesting question of how $\phi $ changes with time as a result
of a changing network topology (in response to the feedback from $\kappa
\left( \phi \right) $), we focus on the steady states after the system
settles down.

\begin{figure}[tbp]
\centering
\begin{overpic}[scale=.4, tics=5,unit=1mm]{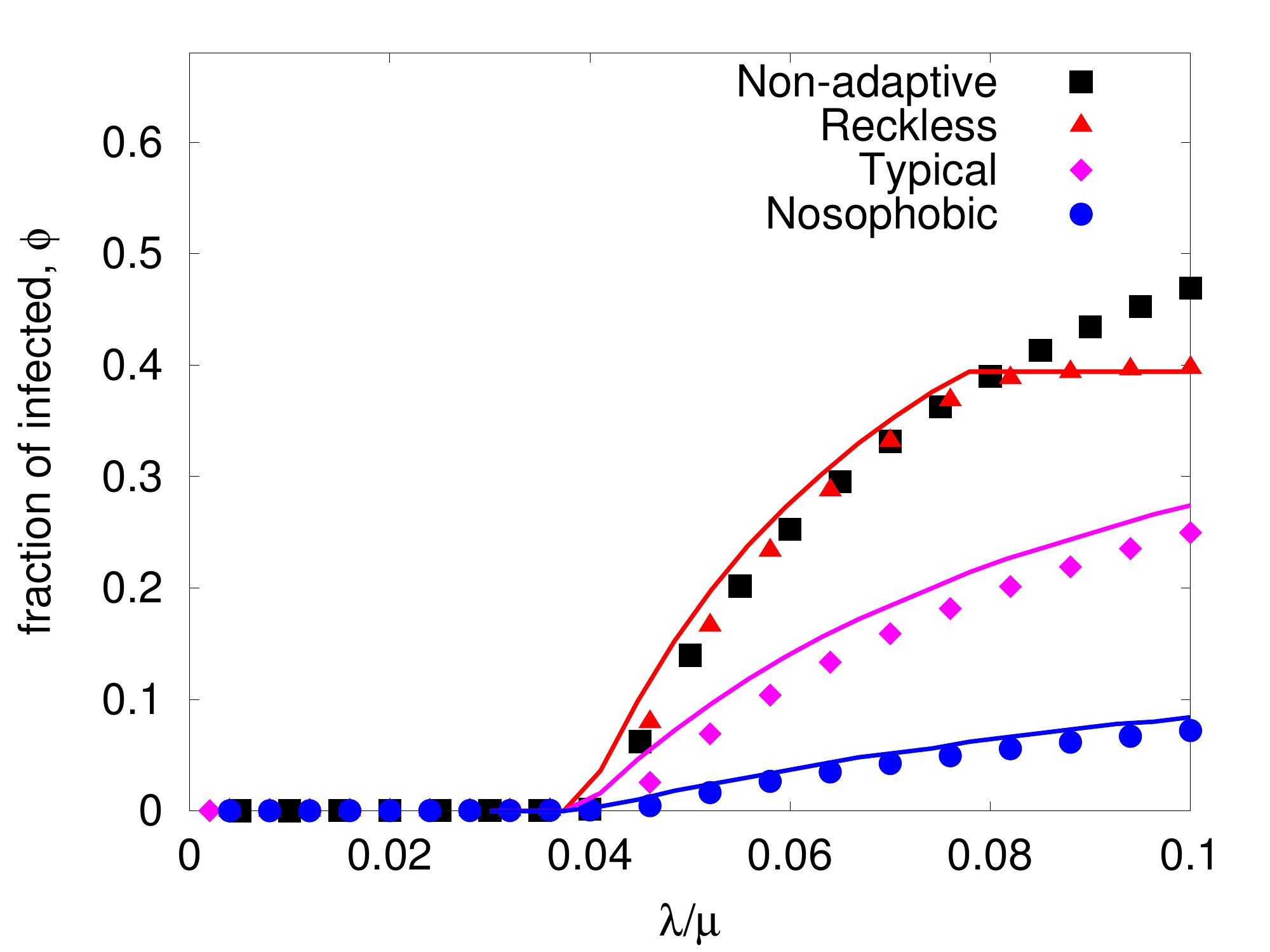}
\put(10,28){\includegraphics[scale=.2]{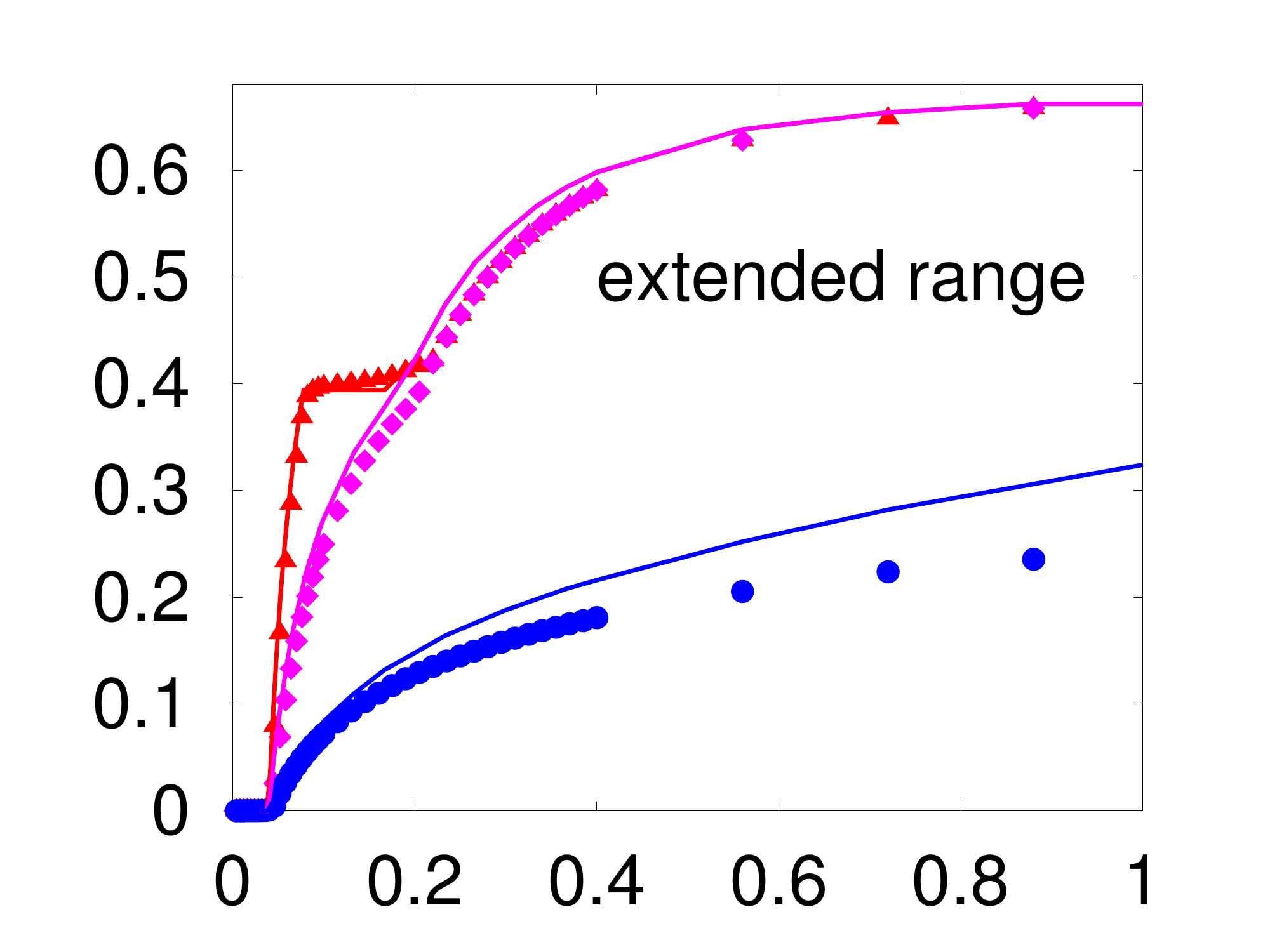}}
\end{overpic} 
\caption{[Color online] {\bf Non-adaptive  and adaptive preferred
degree SIS phase diagram.}  We have chosen $N=5000$, $\kappa_0=25$ and 
$\kappa_{min}=10$ for all three adaptive models (See Figure. 
\ref{Fig3}). The solid lines represent the mean field solution to these
models based on Eq. \ref{phi-lambda-a}. }
\label{Fig4}
\end{figure}

In Figure. \ref{Fig4}, we show the simulation results for $\phi \left( \lambda
\right) $ in these three cases (with mostly $\xi =1$, flexible individuals,
for simplicity), as well as the case above: a non-adaptive network. 
 We first observe that the epidemic thresholds are essentially unchanged by any of the
adaptive strategies. This fact is understandable, since the threshold is
defined by $\phi $ rising from zero and our transition is continuous. Thus,
fear\ in the population has yet to take hold, and $\kappa $ remains close to 
$\kappa _{0}$. Beyond the threshold, the effects of the different fear
factors are self-evident. The reckless follow the non-adaptive until $\phi $
reaches $\phi _{\theta }$ (chosen to be 0.4 here), and then abruptly adjust
their response so that the infection remains more or less at this level. In
the inset, we see that $\phi $ resumes its upward trend after $\lambda/\mu
\simeq 0.2$, and reaches close to the maximal level $\phi _{\max }=2/3$ by $%
\lambda/\mu \simeq 1.0$. By contrast, the infection level in the typical case
increases at a slower pace immediately after $\lambda _{c}$. Around $\lambda/\mu
\simeq 0.2$, $\phi _{typical}\left( \lambda \right) $ coincides with the
reckless, since both networks are controlled by the same $\kappa _{min}=10$.
Finally, as expected, infections in a nosophobic population are strongly
suppressed. Indeed, the critical properties near the transition may be
altered. Since $\kappa \left( \phi \right) $ is effectively zero for $\phi
\gtrsim \phi _{s}\ln \kappa _{0}$ (i.e., $0.1\ln 25\simeq 0.35$ here), it is
not surprising that the infection levels are far lower than the other two
types.

More quantitatively, simple MF theory should provide an acceptable
explanation for these results. From the analysis above, a $\kappa \left(
\phi \right) $ can be readily incorporated, so that $\lambda _{c}^{MF}$
remains unchanged: $1-e^{-\mu /\kappa _{0}}$ . Above this value, the only
modification is the $\lambda $-$\phi $ relationship, and Eqn. (\ref%
{phi-lambda}) now reads%
\begin{equation}
\lambda =1-\left\{ \frac{1-\left( 1+\mu \right) \phi ^{MF}}{1-\phi ^{MF}}%
\right\} ^{1/(\kappa _{0}\phi ^{MF}f\left( \phi ^{MF}\right) )}\,\,.
\label{phi-lambda-a}
\end{equation}%
Although the fear factor appears explicitly here, this expression is quite
cumbersome. A simple way to regard the effects of adaptation is the
following: To produce the same level of infection ($\phi $), the infection
rate ($\lambda $) must be enhanced over the non-adaptive
population. Quantitatively, $-\ln \left( 1-\lambda \right) $ ($\cong \lambda 
$, for small $\lambda $ such as in our examples) must increase by a factor
of $1/f\left( \phi \right) $. In this way, it is easy to see that the MF
prediction of the critical exponent $\beta $ will remain unchanged, unless $%
f $ is appropriately non-analytic at $\phi =0$ (i.e., $\beta =\beta ^{\prime
}$ if $1-f\propto \phi ^{\beta ^{\prime }}$ with $\beta ^{\prime }<1$). At
the other extreme, the saturation levels are given by setting the left side
of Eqn. (\ref{phi-lambda-a}) to unity. Unless the fear factor is so intense
that $f$ vanishes at a value of $\phi $ less than $1/\left( 1+\mu \right) $,
then, strictly speaking, these do not depend on the details of the adaptive
strategy $f\left( \phi \right) $. However, for the severely fearful such as
the nosophobic, the infection essentially levels off at a $\phi $
considerably lower than $\phi _{\max }$. 

Comparing with simulation data, we see that the MF predictions (Figure. \ref{Fig4}) tend to lie a little above simulation data, with the exception of
few points near region associated with the abrupt drop in $\kappa \left(
\phi \right) $ for the reckless population. We believe this effect may be the
result of large fluctuations in the degree distribution. Individuals caught
in this regime may cut ties drastically (at the news of $\phi $ rising above 
$\phi _{\theta }$), causing the infection to decline. But this good news
would lead to the population reversing course just as abruptly, so that
large fluctuations should continue. To test this conjecture, we now present
degree distributions as an indication of how serious these fluctuations can
be.

\begin{figure*}[tbp]
\centering 
 \includegraphics[scale=1.0,trim=4.2cm 12.0cm 0.5cm 5.5cm,clip]{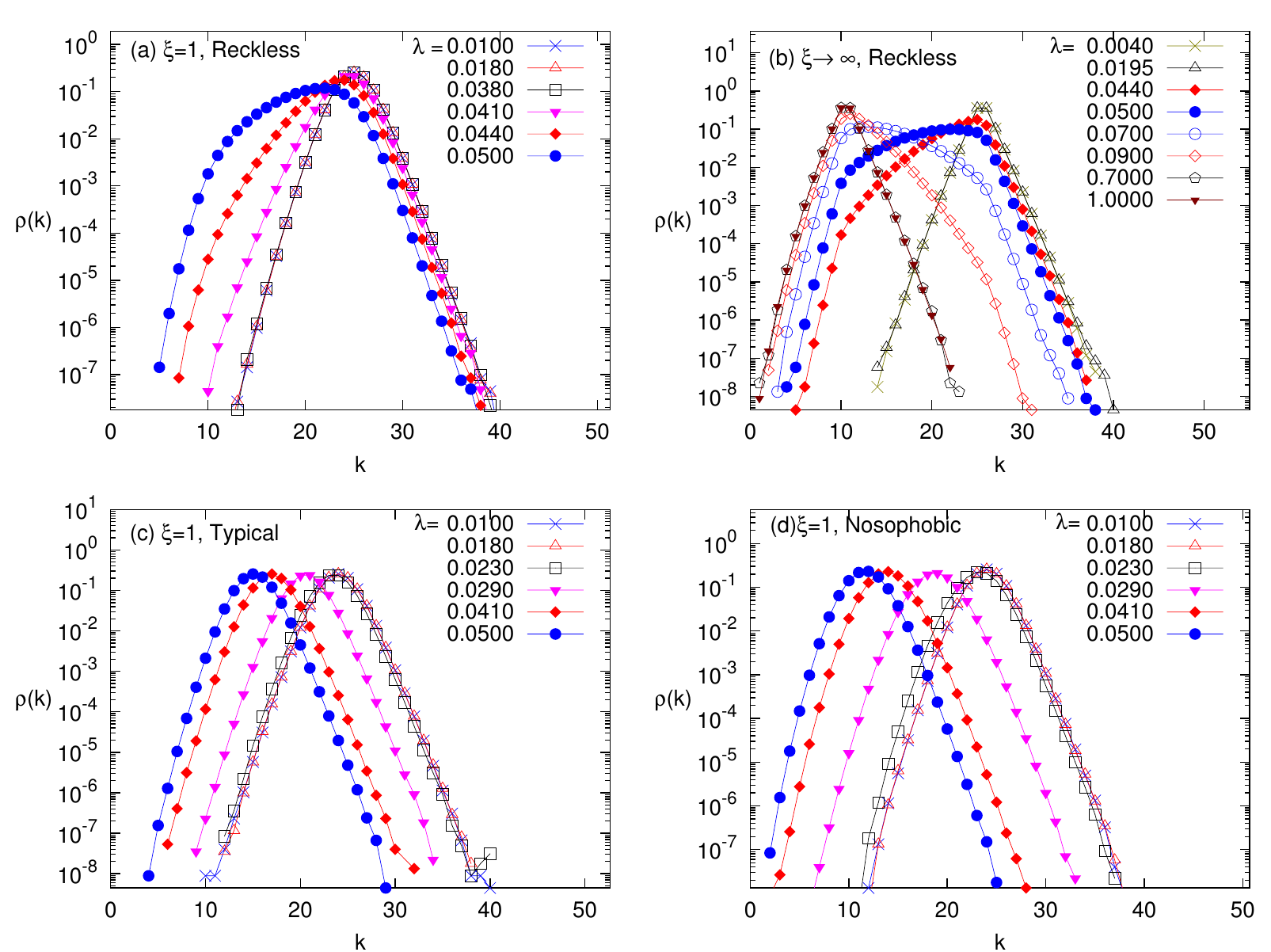}
\caption{[Color online] {\bf Steady state degree distribution of adaptive network. }
Degree distribution of (a) reckless with $\xi%
=1$ (see Eq. \ref{plink}) (b) reckless and inflexible individuals ($%
\xi=\infty$), (c) Typical and (d) Nosophobic individuals (see Sec
3.A for details) with $\xi=1$. We have chosen $N=5000, \kappa%
_0=25, \kappa_{min}=10$ for all these cases. The infection rates $%
\lambda$ are chosen to illustrate transition behavior in degree
distributions.}
\label{Fig5}
\end{figure*}

In the absence of infection, the degree distribution should be similar to
those in Figure.\ref{Fig1}, around the preferred $\kappa _{0}$. Far from the
transition, the epidemic has settled in and, for both the typical and the
reckless, the distribution should also be similar, but settling around $%
\kappa _{min}$ instead ( Figure.\ref{Fig5} ). Not surprisingly, the picture is more complex for
the nosophobic, especially for large $\lambda $, since the preferred degree
is strongly dependent on the level of the infection and approach zero, which tends to isolating the nodes. Here, let us focus on the effects of the abrupt behavior of the reckless, the case that also
displays the most interesting behavior (large fluctuations, Figure.\ref{Fig5}
a,b). For the other two types, we note the predictably mild changes in the
degree distribution, as $\lambda $ increases (Figure.\ref{Fig5}c,d). The
overall \textit{shape} of $\rho\left( k\right) $ remaining essentially the
same, but due to adaptations the center slowly shifts with $\phi $ and $%
\lambda $.

For the reckless population, the conjectured behavior --dramatic swings when
the infection level is near $\phi _{\theta }$, is well captured in the
broadening of $\rho \left( k\right) $. From the data shown in Figure.\ref{Fig5}%
a, we see that the distributions are, as expected, centered close to $\kappa
_{0}$ for $\lambda \lesssim 0.04$ ($\lambda \thicksim 0.04$ corresponding to
the threshold $\phi _{\theta }\thicksim 0.4$). Thereafter, many individuals
in the population begin to cut contacts. By $\lambda =0.05$, $\rho \left(
k\right) $ is quite distorted compared to the simple Laplace distribution.
Specifically, we see that a sizable fraction of the population has cut their
preferences down towards $\kappa _{min}=10$. To display a complete range of
infection rates, we chose to simulate with rigid individuals ($\xi =\infty $%
) for simplicity (Figure.\ref{Fig5}b). Here, we see the complete crossover as $%
\lambda $ increases, from a distribution centered around $\kappa _{0}$ to
one around $\kappa _{min}$. If we plot a reflected and
appropriately shifted version of the $\lambda =0.065$ distribution (i.e., $%
\rho (\tilde{k}-k) $ for an appropriate $\tilde{k}$), the result
is essentially identical to the raw $\rho \left( k\right) $ for $\lambda
=0.050$. A similar collapse is observed for the cases with $\lambda =0.055$
and $\lambda =0.060$, $\rho \left( k\right) $. Thus, we may associate $%
\lambda _{f\text{-}t}\cong 0.057$ with a transition, from a\ population
dominated by $\kappa _{0}$ (i.e., non-adaptive behavior) to one controlled by $%
\kappa _{min}$ (i.e., typical ). Since $\rho ^{\ast }\left( k\right) $
displays always a single peak, which shifted rapidly between $\kappa _{min}$
and $\kappa _{0}$, we would label this as a continuous transition. 

\subsection*{III.c ~~ Selective adaptation}

\begin{figure*}[htbp]
\centering
\includegraphics[scale=1.0,trim=4.0cm 10.2cm 2.5cm 5.6cm,clip]{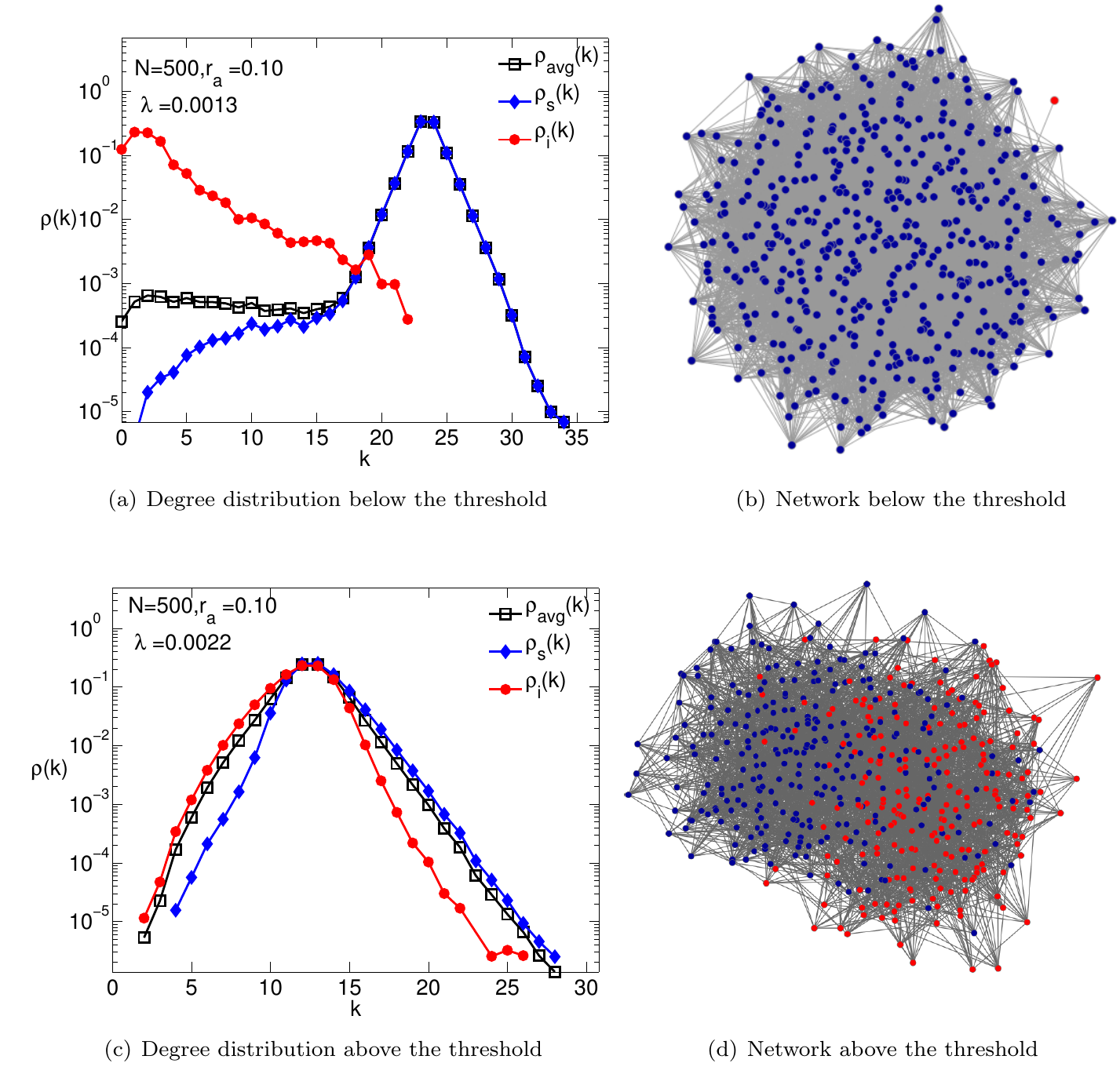}
\caption{[Color Online] {\bf Degree distribution and network structures with
typical local adaptations. }  Panels (a) and (b) show systems below the epidemic threshold, while (c) and (d) show systems above the threshold. The parameters chosen are $N=500, \kappa_0=25,\kappa_{min}=10, \gamma=10\kappa_0, r_a=0.1$, $\lambda/\mu=0.1 $. }
\label{LocalBTr}
\end{figure*}

If the state of infected individuals is manifest (i.e., disease is
`visible'), it is natural for individuals to be more selective in
choosing their contacts. Such behavior might also be driven by policy
interventions such as isolating the infected and/or closing public meeting
grounds (e.g., schools) \cite{Hatchett.PNAS.2007, SARS.2003}. 
In particular, how an individual adds/cuts links will now depend on the
states of his/her contacts. We choose the following `think globally, act
locally' model which we believe is a reasonable representation of such
adaptive behavior.

We initially set up a static preferred degree network with a preferred
degree $\kappa =\kappa _{0}$. Infection is started in some fraction $\phi_{0}$
of the nodes and spreads according to the standard SIS dynamic rules
described before. As in the blind adaptation case, the preferred degree $\kappa
=\kappa _{0}f(\phi )$ depends on the global infection level $\phi $. Unlike
the previous method, when a node is chosen to update its links, the rules
will depend on whether the node is susceptible or infected. Let us assume
that an $I$ does not care about the state of the contacts and randomly
adds/cuts links as before. However, an $S$ will  behave more \textit{selectively}, having a bias in favor of other $S$'s after it decides to add
or cut a link. To model this bias, we introduce a parameter, $\gamma $, with which the favored choice is selected over the undesirable one. Letting subscripts denote the initiator-receptor pairs, $p_{SI}$ and $p_{SS}$ denote, respectively, the probability with which an $S$ cuts a link to an $I$ or an $S$. Obviously, we impose $p_{SI}+$ $p_{SS}=1$. Similarly, let $\tilde{p}_{SI}$ and $\tilde{p}_{SS}$ denote the probabilities it will \textit{create}, respectively, a
link to an $I$ and an $S$ (with $\tilde{p}_{SI}+\tilde{p}_{SS}=1$). Explicitly, we choose the following.
\begin{itemize}
\item An $S$ with degree $k\geq \kappa $ will \textit{cut} a link from a
randomly chosen $I$ with probability 
\begin{equation}
p_{SI}=\frac{\gamma k_{I}}{\gamma k_{I}+k_{S}},  \label{PSI}
\end{equation}%
or to a randomly chosen susceptible with probability $p_{SS}=1-p_{SI}$.
Here $k_{I},k_{S}$ are the number of $I,S$ contacts it has. Now, it is clear
that the larger $\gamma $ is, the more our $S$ will choose to cut links to its infected
contacts ($\gamma =1$ corresponds to non-preferential adaptation).

\item Similarly, an $S$ with degree $k<\kappa $ will \textit{create} a link
to a randomly chosen $S$ with probability 
\begin{equation}
\tilde{p}_{SS}=\frac{\gamma k_{S}}{\gamma k_{S}+k_{I}},  \label{PTSS}
\end{equation}%
or to a randomly chosen infected with probability $\tilde{p}_{SI}=1-\tilde{p}%
_{SS}$. Again, we see a large $\gamma $ biases more towards adding links to
other $S$'s. 

\item Since infected nodes do not have any incentive for selective
adaptation, we make these nodes adapt blindly as follows:
\begin{equation}
p_{II}=~\tilde{p}_{II}=\frac{k_{I}}{k_{I}+k_{S}};~~p_{IS}=\tilde{p}_{IS}=%
\frac{k_{S}}{k_{I}+k_{S}}~.
\end{equation}
\end{itemize}
To allow for individuals to react at a different rate compared to
that of recovery or infection, as in blind adaptation case, we update the links  at
a rate $r_{a}$ ($<1$) compared to the update of the state of the nodes.

\begin{figure*}
 \setlength{\unitlength}{0.1\textwidth}
\centering
\includegraphics[scale=1.0,trim=5.5cm 16.6cm 1cm 0.60cm,clip]{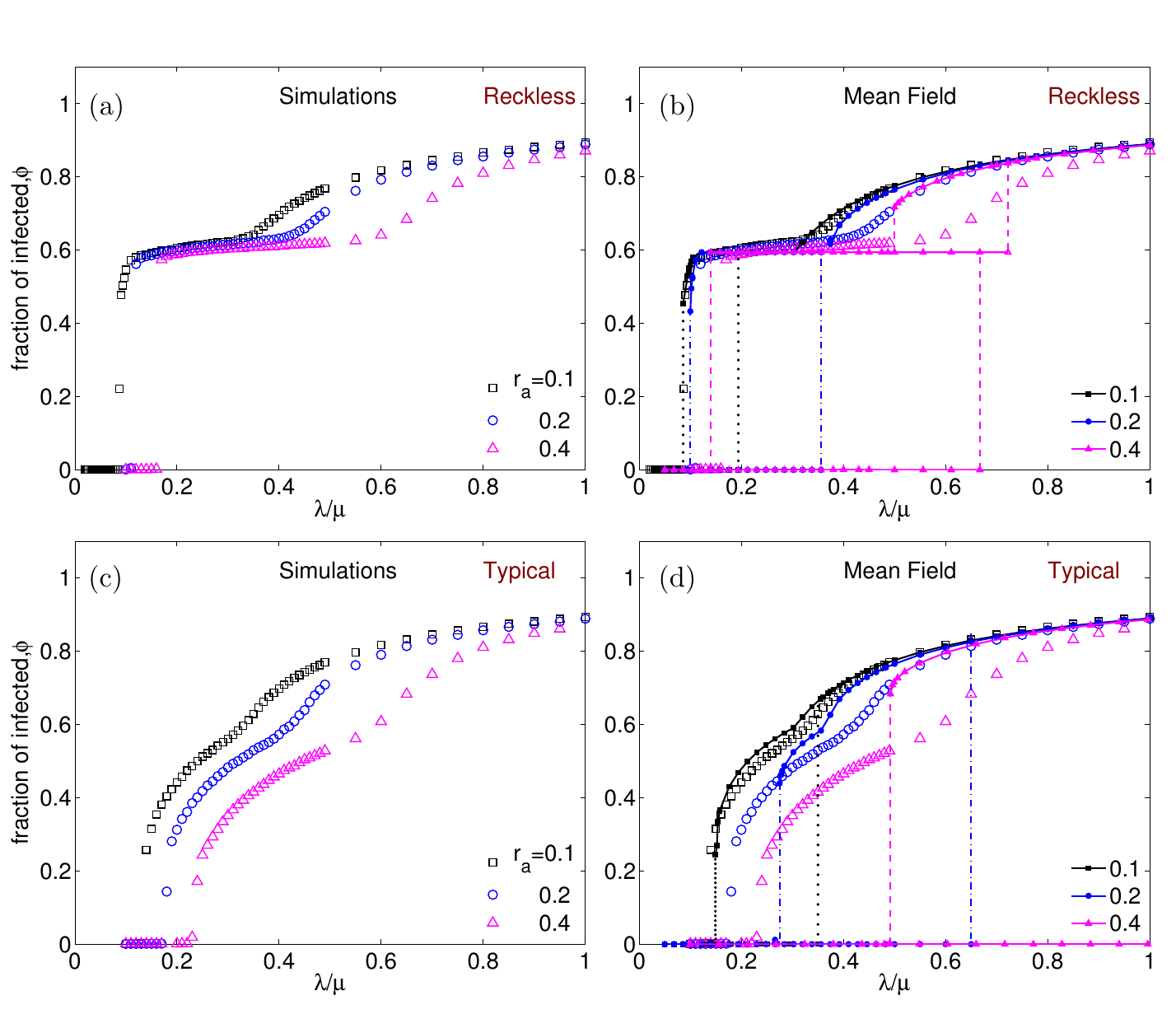}
\caption{[Color Online] {\bf SIS phase diagram for selective adaptations. } The fraction of infected  population $\phi$, versus $\lambda/\mu$ for different network adaptation rates, with parameters $N=1000$, $\kappa_0=25$, $\kappa_{min}=10$, $\gamma/\kappa_0=10$. Panels (a) and (c) show the Monte-Carlo simulation results for reckless and typical behaviors (see Sec. III.A) respectively. In panels (b) and (d), the simulation results are compared to local mean field theory (described in Appendix) predictions. The black squares, blue  circles and magenta triangles represents the network adaptation rates $r_a=0.1,0.2$ and $0.4$ respectively. The corresponding mean fields results are plotted as lines with respective colors in (b) and (d). The dotted, dot-dash and dashed lines represent the bistable regions obtained from mean field  solutions when initial infection fraction is varied from $\phi_0=0.05~\mathrm{to}~0.8$ and initial links chosen from  following the hysteresis curve. }
\label{LocalPhDia}
\end{figure*}

With the rules described, we studied selective adaptations for reckless
and typical cases (see section. II.a ) for moderate system sizes $N=500,1000$.
We found that system size satisfying $N > \kappa_{max}^2$ is sufficient to
produce the `thermodynamic' limit. While we note that steady state
configuration depend only on the ratio $\lambda/\mu$, we alert the readers
that our parameters $\mu,\phi_\theta$ for selective adaptation are different
from the blind adaptation case. We choose $\mu=0.01\ll \mu_{blind}=0.5$, and
the cut off infection level for $\kappa=\kappa_{min}$ to be 60\% of the
maximum value $\phi_\theta=0.6\phi_{max}=0.6/(1+\mu)\approx 0.6$.

In Figure. \ref{LocalBTr}a, we show the degree distribution of susceptibles,
infected and total populations \textit{below} the epidemic threshold for a
typical behavioral adaptation case. Except for one immortal, the whole
population is composed of susceptibles. The total degree distribution
essentially reflects the susceptibles. However, the immortal can have
different degrees during the course of SIS dynamics which will be reflected
in the quenched distribution of infected. Figure. \ref{LocalBTr}b shows the
network structure with the lone infected connected to the big cluster of
susceptibles. In Figure. \ref{LocalBTr}c, we show the degree distribution of
susceptibles, infected and total populations \textit{above} the epidemic
threshold with parameters $r_a=0.1$ and $\lambda/\mu=0.22$. We see that all
the degree distributions overlap. However the infected people are more
strongly interconnected than with the susceptibles (see Figure. \ref{LocalBTr}d), which is
indicated by non-zero modularity coefficient \cite{AaronNewman.PRE.2004,Newman.PNAS.2006} of Q=0.2384. 

In Figure. \ref{LocalPhDia}a and c, we show the SIS phase diagram for reckless
and typical adaptations obtained by Monte-Carlo simulations. In the figure,
black squares, blue circles and magenta triangles correspond to relative
network adaptation rates $r_{a}=0.1,0.2,0.4$ respectively. We observe that
unlike the blind adaptation case, the epidemic threshold varies both with
the network adaptation rate and behavioral response to different fear levels.
The threshold increases with increasing $r_{a}$ -- an understandable feature, as
faster responses by the $S$'s should suppress the infection rates. In both
cases, the transition from a healthy state to an active infectious state is
considerably more rapid than in the blind adaption case. Indeed, for the
reckless population with faster network response (larger $r_{a}$), we
observe a discontinuous transition (or a very steeply rising continuous
one). In both cases, there is a second crossover, near $\phi \thickapprox
0.6 $, to a gently rising $\phi \left( \lambda /\mu \right) $ curve. These
can be traced to our choice of $\kappa \left( \phi \right) $, which
contains a singularity (discontinuity or kink) at $\phi _{\theta }\left(
\cong 0.6\right) $.

Since the adaptation is in response to a `local' environment of a
susceptible individual, a more sophisticated mean field theory needs to be
formulated. To distinguish this from the mean field approach above, we will
refer to it as the `local mean field theory' (LMFT). In particular, we
introduce three more variables: $l_{SI},l_{SS}$, and $l_{II}$, defined as
the mean number of $SI,SS$ and $II$ links per node, respectively. While the
evolution equation for $\phi $ is just modified to be $d\phi /dt=-\mu \phi
+\lambda l_{SI}$, the equations for the $l$'s are much more involved.
Deferring to the Appendix (see supplementary information ) the details of how these are formulated and
studied, let us focus here on the results of the stationary solutions, Eqns.
(13, 20) of Appendix, and how they compare with simulation data.
Illustrated in Figure. \ref{LocalPhDia}b and d, the general conclusion is that
there is reasonable qualitative agreement between LMFT and Monte Carlo
results.

For the case with reckless adaptations, the response to infections is quite
rich while the agreement is better than expected. In particular, LMFT
predicts \textit{three} stable fixed points: one associated with the
inactive $\phi =0$, another associated with $\phi _{\theta }$, and the
third, with a `normal' endemic state. The presence of the second fixed point
is probably the result of the discontinuity in our $\kappa _{reckless}(\phi
) $. Moreover, for a moderate range of $\lambda $, the LMFT displays
bistability. Of course, in a stochastic simulation, one of these will be
metastable with a discontinuous transition in $\phi \left( \lambda \right) $.
Such differences are common, much like bistability in a Landau theory of
ferromagnetism below criticality \textit{vs.} metastability/stability in a
statistical system. Overall, we see that simulation data generally support
the existence of \textit{three} branches, in good agreement with LMFT. In
more detail, we find that the nature of the first transition (threshold of
the epidemic, from the inactive state to $\phi \cong \phi _{\theta }$) is
well predicted by LMFT. Comparing the location of the discontinuous
transition is, of course, very difficult. Nevertheless, simulations indicate
these locations to lie within the LMFT limits of bistability. In any case,
there is good reason to believe that the (bare) value of $r_{a}$ (from
simulations) will be `renormalized' by fluctuations, so that a better theory
may converge towards the data. Turning to the second transition, at higher $%
\lambda $, we see that it is associated with $\phi $ exceeding $\phi
_{\theta }$, which in turn leads to a jump in $\kappa $ (from $\kappa _{0}$
to $\kappa _{min}$). Thus, the network will become homogeneous again: With
degree $\kappa _{min}$, the theoretical $\phi \left( \lambda \right) $ follows $\lambda \kappa _{min}/\left( \lambda \kappa_{min}+\mu \right) $. This prediction agrees with simulations, once $\lambda 
$ far exceeds the transition values. More intriguingly, LMFT predicts the
nature of this transition to depend on $r_{a}$. While it is a typical
bifurcation for the lower $r_{a}$'s, it a involves {\it tri-stability} region ($\lambda/\mu=0.5-0.65$), with  all the three branches are stable for the $r_{a}=0.4$ case. In the latter case, the LMFT displays oscillating time
dependence in all the variables in the $\phi=\phi_\theta$ branch, pointing to the possibility of limit cycles and Hopf bifurcations. Perhaps just an artifact of the discontinuity in $\kappa _{reckless}\left( \phi \right) $, these fascinating aspects deserve
further study. Comparisons with data are more ambiguous. For example,
simulations favor gentle crossovers rather than discontinuities in $\phi
\left( \lambda \right) $ or $d\phi /d\lambda $. Remarkably, the location of
these crossover are not too far from the transition predicted by the LMFT.

For the `typical' adaptive behavior, we find two stable fixed points corresponding to the inactive or endemic
states. Moreover, for a moderate range of $\lambda $, the LMFT displays
bistability, i.e., it predicts a discontinuous transition. The agreement
between LMFT and simulation results is arguably good for $r_{a}=0.1$,
finding even the kink associated with $\kappa _{typical}(\phi )$ at $\phi
_{\theta }$. For larger $\lambda/\mu$, the branch of the LMFT bistable region and the data follows $\lambda \kappa _{min}/\left( \lambda \kappa_{min}+\mu \right)$ for all $r_a$'s.  For the larger $r_{a}$'s, the theory continues to predict a discontinuous transition at the threshold, while the data show a steadily decreasing discontinuity. It is quite possible that these end on a multicritical point, beyond which the behavior is more typical of a `second order' transition. Such subtle issues can only be clarified with a larger systematic simulation study.   The reasons for the discrepancy between LMFT and simulations are unclear. We speculate that some of the approximations used were too crude, e.g., replacing the local degrees with the global averages (see Appendix in supplementary information for details) and assuming degree distributions to adopt instantaneously to the steady state adaptive preferred degree (with a time dependent $\kappa $). These are issues worthy of further investigation. Clearly, there 
is considerable room for 
improvement as many questions remain to be explored before we arrive at a satisfactory theory.

\section*{Conclusions}

The study of dynamical processes on networks has been very active for several
decades. Most investigations have focused on either a dynamic set of nodes
on a static network (e.g., spins on a lattice or epidemics in a population
with fixed connections) or a dynamic network with static nodes (e.g., small
world networks, scale free networks).  Only recently have researchers focused their attention on dynamics of co-evolving networks where both nodes and links are dynamic, with particular attention to opinion dynamics and epidemic
spreading. Here we consider the classic SIS model of epidemic spreading, on
a network that adapts to the level of the infection. Introducing a new class
of networks in which individuals (nodes) favor a certain number of contacts (%
$\kappa $, the preferred degree), we model various types of adaptive
behavior by letting $\kappa $ depend on the level of the epidemic, through $%
\phi $, the infected fraction of the population. For such networks, we
typically find degree distributions that are neither Gaussian nor
scale-free. Instead, the universal feature appears to be exponential tails
when the degree is far from $\kappa $.

Using Monte-Carlo methods, we simulated populations in which healthy
individuals may become ill by being in contact with a fluctuating set of
infected nodes, while diseased persons recover spontaneously with some rate.
We considered three types of adaptive behavior representing the degree of
fear in the public, which were modeled by different adaptive preferred
degree as a function of global infection level.  Further, these network adaptations can be \emph{blind}, i.e., a central node does not know the disease state of its contacts, or \emph{selective} where the disease state of the neighbors is known and the central node responds by selectively cutting or creating links.   For
the blind adaptations we find that the epidemic threshold does not change with
the degree of fear, however the level of epidemic in the active phase decreases
with increasing fearful response. A good agreement with the simulation data
can typically be found with a simple mean field theory. For the selective
adaptations, much more interesting dynamics emerge. The epidemic threshold
changes substantially with increasing rate of network adaptations ($%
r_{a} $). The epidemic transition is discontinuous, unlike the blind
adaptation case which shows a continuous transition. The level of epidemic in
the active phase changes with both the network adaptation rate and the
degree of fear in the public. We have presented a local mean field theory
with equations for both node and link dynamics for selective adaptations.
For reckless and typical cases, it predicts bistable regions in which both, a healthy and an active infectious phase persist - a standard indicator of discontinuous transitions. There is
qualitatively good agreement between mean field predictions and simulation
data. Sources for the (quantitative) differences abound, from the crude
level of approximations used to the subtle effects of fluctuations.

Within the scope of our study, many issues remain to be investigated and
better understood. Clearly, our mean field treatment relied on significant approximations; how can this approach be improved?  Do the observed discontinuous
transitions share typical aspects of `first-order' transitions, e.g.,
hysteresis and metastability? If so, does our system fall into the universality class of the standard SIS problem? Are there new exponents, associated with the network fluctuations and its dynamics?  
At a more detailed level,
insights into much of the properties of the network (e.g., degree
distributions, clustering, modularity, etc.), especially in the case with
selective adaptations, would be very desirable. 

Apart from the two types of adaptation we have presented, many extensions can be pursued. In a typical
society, the population is inhomogeneous, so that an individual's perception
of the infection level may not be the same as the overall $\phi $. Letting
the adaptive behavior depend on this perceived level, we consider variations
in strategies by simply adding a white noise to $\phi $. Our preliminary
studies with `blind' adaptations, not reported above, indicate that the
effect of this type of noise on the epidemic appears to be minimal. Beyond our simple model, the most immediate generalization is to include
spatial structures, both homogeneous and heterogeneous. For example,
extroverts and introverts have very different preferred degrees. How does an
epidemic develop across these different communities? There is a general
belief that extroverts are more prone to contagious diseases.  A further generalization would be to study epidemics on realistic networks with known degree distributions and clustering.  Such networks can be synthesized by heterogeneous preferred degree networks with appropriate built through 
`small world' algorithms. We postpone such work to a future publication. 
Naturally, the long term interest in such studies is to develop a good
understanding so that reasonable public policies can be formulated in
response to a real epidemic.

\section*{Acknowledgements}

We thank Stephen Eubank, Thierry Platini, Leah Shaw and Max Shkarayev for
illuminating discussions. 

\section{Appendix: Local mean field theory for selective adaptation}

Here we discuss a formalism to understand the dynamics of global infection
level and links for selective adaptation. We define the `spin variable' $%
\sigma _{n}=\{0,1\}$ to denote the susceptibles (0) and infected (1)
individuals of node $n~(=1,2\dots ,N)$. The fraction of infected
individuals for any configuration is $\sum_{n}\sigma _{n}/N$. In the mean
field approach, only its average is kept, so we simply replace
$\frac{1}{N}\langle\sum_{n} \sigma _{n} \rangle \rightarrow \phi.$

Meanwhile, a network is uniquely specified by the adjacency matrix 
$a_{mn}=\{0,1\}$, representing the absence or presence of the (undirected)
link between nodes $m$ and $n$.  The degree of node $n$ is then given by $k_{n}=\Sigma
_{m}a_{nm}$ so that the familiar degree distribution is given by $\rho
\left( k\right) =\Sigma _{n}\delta (k_{n}-k)/N$. As we have two kinds of
nodes, let us define two separate degree distributions, for the $S$'s and
the $I$'s: 
\begin{eqnarray}
\rho _{S}(k) &=&\frac{1}{N\left( 1-\phi \right) }\sum_{n}(1-\sigma
_{n})\delta (k_{n}-k)  \nonumber \\
\rho _{I}(k) &=&\frac{1}{N\phi }\sum_{n}\sigma _{n}\delta (k_{n}-k).
\end{eqnarray}%
Note that each is normalized, so that $\rho \left( k\right) =\left( 1-\phi
\right) \rho _{S}\left( k\right) +\phi \rho _{I}\left( k\right) $. Observe
that the average of $\sum_{m,n}a_{mn}/2N$ is $\langle k\rangle /2$, which
is, in a network with preferred degree, just $\kappa /2$. 

Let $l_{SI},l_{SS},l_{II}$ denote these averages \textit{per node}, respectively: 
\begin{eqnarray}
l_{SI} &=&\frac{1}{2N}\sum_{m,n}\left[ (1-\sigma _{m})a_{mn}\sigma
_{n}+\sigma _{m}a_{mn}(1-\sigma _{n})\right]   \nonumber \\
l_{SS} &=&\frac{1}{2N}\sum_{m,n}\sigma _{m}a_{mn}\sigma _{n}  \nonumber \\
l_{II} &=&\frac{1}{2N}\sum_{m,n}(1-\sigma _{m})a_{mn}(1-\sigma _{n}).
\end{eqnarray}%
Thus, we should have 
\begin{equation}
l_{SI}+l_{SS}+l_{II}=\langle k\rangle /2.  \label{Sum-l-kappa}
\end{equation}
Turning to \textit{dynamics}, the equation for $\phi $ is given, in the mean
field approximation, by: 
\begin{equation}
\frac{d\phi }{dt}=-\mu \phi +\lambda l_{SI}.  \label{IEqn}
\end{equation}%
The dynamical equations for these links are more involved. For ease of
understanding, we split the link equations into three parts, separating
effects of node dynamics (infection, recovery) and network adaptations \cite%
{ZanetteR.JBioPhys.2008}.

\paragraph{\textbf{Node Dynamics:}}

When the network topology is fixed, but the state of nodes are changing due
to infection and recovery process, equations for the links can be written
as: 
\begin{subeqnarray}
\frac{dl_{SI}}{dt} &=&\lambda \left[ 2 \frac{l_{SS}l_{SI}}{1-\phi}- \frac{l_{SI}^2}{1-\phi}- l_{SI}\right]  -\mu \left [l_{SI}+2 l_{II} \right]\\ 
\frac{dl_{SS}}{dt} &=&-2\lambda \frac{l_{SS}l_{SI}}{1-\phi}+ \mu l_{IS}  \\ 
\frac{dl_{II}}{dt} &=&\lambda \left[\frac{l_{SI}^2}{1-\phi}+ l_{SI}\right] -2\mu l_{II}.
\label{InfAdEq}
\end{subeqnarray}Here we have used the standard moment closure approximation
for triplets $l_{abc}=l_{ab}l_{bc}/l_{b}$, with $a,b,c$ being $S$ or $I$. 
\cite{GrossDLimaB.PRL.2006,Keeling.JRSi.2005}. The term $l_{SSI}\approx 
\frac{l_{SI}l_{SS}}{1-\phi }$ in Eq. \ref{InfAdEq}a (b) corresponds increase (decrease) 
$SI$ ($SS$) links due to [$SSI$] triplet. $l_{ISI}\approx 
\frac{l_{SI}^{2}}{1-\phi }$ corresponds to decrease in $SI$ (increase in $II$%
) link. In both infection and recovery process, the total number of links is
conserved.

\paragraph{\textbf{Network Adaptations}:}

But, our links are also being created and cut, at a rate $r_{a}$ relative to
the node dynamics, according to the rules of selective adaptation described
in section III.C. Thus, we must add such terms to the $dl/dt$ equations.
Within the spirit of mean field theory, these are given by 
\begin{eqnarray}
\frac{1}{r_{a}}\frac{dl_{SI}}{dt} &=&\sum_{k}(1-\phi )\rho _{S}(k)\left[
\Theta (\kappa -k)\tilde{p}_{SI}-\Theta (k-\kappa )p_{SI}\right]  +\sum_{k}\phi \rho _{I}(k)\left[ \Theta (\kappa -k)\tilde{p}_{IS}-\Theta
(k-\kappa )p_{IS}\right]   \nonumber \\
\frac{1}{r_{a}}\frac{dl_{SS}}{dt} &=&\sum_{k}(1-\phi )\rho _{S}(k)\left[
\Theta (\kappa -k)\tilde{p}_{SS}-\Theta (k-\kappa )p_{SS}\right]   \nonumber
\\
\frac{1}{r_{a}}\frac{dl_{II}}{dt} &=&\sum_{k}\phi \rho _{I}(k)\left[ \Theta
(\kappa -k)\tilde{p}_{II}-\Theta (k-\kappa )p_{II}\right] .  \label{NWAdEq}
\end{eqnarray}%
To simplify, we absorb the sums of $k$ into simplified
expressions: 
\begin{eqnarray}
\Sigma _{S}^{-} &\equiv &\sum_{k}\rho _{S}(k)\Theta (k-\kappa )(-1), 
\nonumber \\
\Sigma _{S}^{+} &\equiv &\sum_{k}\rho _{S}(k)\Theta (\kappa -k)(+1), 
\nonumber \\
\Sigma _{S} &\equiv &\Sigma _{S}^{+}+\Sigma _{S}^{-}=\sum_{k}\rho _{S}(k)%
\mathrm{sgn}(\kappa -k),  \nonumber \\
\Sigma _{I} &\equiv &\sum_{k}\rho _{I}(k)\mathrm{sgn}(\kappa -k).
\label{SumRhos}
\end{eqnarray}
To continue, we approximate the local degrees of the susceptibles by the
global averages. For the susceptibles we replace $(k_{S},k_{I})\rightarrow
(l_{SS},l_{SI})$ so that the probabilities in Eq.\ref{PSI},\ref{PTSS} become independent of $k$%
\begin{equation}
p_{SI} =1-p_{SS}\rightarrow \frac{\gamma l_{SI}}{\gamma l_{SI}+l_{SS}} ; \quad 
\tilde{p}_{SS} =1-\tilde{p}_{SI}\rightarrow \frac{\gamma l_{SS}}{\gamma
l_{SS}+l_{SI}}. 
\end{equation} 
Similarly, for the infected, we use $k_{S}\rightarrow
l_{IS}(=l_{SI}),~k_{I}\rightarrow l_{II}$ so that%
\begin{equation}
p_{IS}=1-p_{II}=\tilde{p}_{IS}=1-\tilde{p}_{II}\rightarrow \frac{l_{SI}}{%
l_{SI}+l_{II}}  
\end{equation}

Since the degree distributions also vary with time, the sums in Eq. \ref{SumRhos} cannot
be expressed in terms of the mean field variables ($\phi
,l_{SI},l_{SS},l_{II}$) on which we have chosen to focus. To proceed, we
make a further (drastic) assumption, that each $\rho $ can be approximated
by the Laplacian distribution of Section II.A, around the instantaneous $%
\kappa \left( \phi \right) $. Technically, this assumption gives rise to an
unphysical constraint, namely, \textit{symmetric }degree distributions
conserves the total number of links. But, this contradicts Eqn. (\ref%
{Sum-l-kappa}), since $\kappa (\phi (t))$ cannot be a constant. To ensure
that, during adaptations,  $l_{SI}+l_{SS}+l_{II}=\kappa (\phi )/2$ is
satisfied, we introduce an auxiliary `damping' field:%
\begin{equation}
\eta =-r_{a}(l_{SI}+l_{SS}+l_{II}-\kappa (\phi )/2)
\end{equation}%
into the evolution equations of the $l$'s. With these modifications, the
final set of mean field equations for the links (including recovery,
infection and network adaptation process) read: 
\begin{eqnarray}
\frac{dl_{SI}}{dt} &=&\alpha \eta +r_{a}\left[ (1-\phi )\left( \tilde{p}%
_{SI}\Sigma _{S}^{+}+p_{SI}\Sigma _{S}^{-}\right) +\phi p_{IS}\Sigma _{I}%
\right]   -\mu l_{SI}+2\mu l_{II}+2\lambda \frac{l_{SS}l_{SI}}{1-\phi }-\lambda 
\frac{l_{SI}^{2}}{1-\phi }-\lambda l_{SI}  \nonumber \\
\frac{dl_{SS}}{dt} &=&(1-\alpha -\beta )\eta +r_{a}(1-\phi )\left( \tilde{p}%
_{SS}\Sigma _{S}^{+}+p_{SS}\Sigma _{S}^{-}\right) + \mu l_{SI}-2\lambda \frac{l_{SS}l_{SI}}{1-\phi }  \nonumber \\
\frac{dl_{II}}{dt} &=&\beta \eta +r_{a}\phi p_{II}\Sigma _{I}-2\mu
l_{II}+\lambda \frac{l_{SI}^{2}}{1-\phi }+\lambda l_{SI},  \label{NWAdEqSim}
\end{eqnarray}%

where the `damping coefficients' $\alpha ,\beta $ are somewhat arbitrary.
They must be chosen to model the fact that, as the infection rages, $SI$ and 
$II$ links should decrease while $SS$ links should increase. Thus, we impose 
$\alpha ,\beta >0$ and $\alpha +\beta >1$. The link equations in \ref{NWAdEqSim} along with the node Eqn.\ref{IEqn} and fear function Eq. \ref{fearfunction} forms the set of mean field equations for selective
adaptations.

We evolve the mean field equations numerically and obtain the stationary
state infection and links. The time of evolution varied from 500-1000 units
for reaching steady state. We chose $\alpha =1.0,~\beta =0.5$, $\lambda /\mu
\in \left[ 0,2\right] $ and a range of initial infections so as to find the
various stable fixed points shown in the text. The remaining parameters are the
same as those used for Monte-Carlo simulations.

\bibliographystyle{plos2009}
\bibliography{references}

\end{document}